\documentclass{article}
\usepackage{arxiv}
\usepackage{dcolumn}
\usepackage{mathptmx} 
\usepackage{multirow}
\usepackage{multicol}
\usepackage{rotating}

\usepackage[utf8]{inputenc} 
\usepackage[T1]{fontenc}    
\usepackage{hyperref}       
\usepackage{url}            
\usepackage{booktabs}       
\usepackage{amsfonts}       
\usepackage{nicefrac}       
\usepackage{microtype}      
\usepackage{lipsum}
\usepackage{amssymb,amsmath,epsfig}
\newcommand{\beq}{\begin{equation}}
	\newcommand{\eeq}{\end{equation}}
\newcommand{\bea}{\begin{eqnarray}}
	\newcommand{\eea}{\end{eqnarray}}

\begin{document}
\title{Optical features of rotating quintessential charged black holes in de-Sitter spacetime}
\author{{Saeed Ullah Khan$^1$$^2$\thanks{drkhan@szu.edu.cn, saeedkhan.u@gmail.com}}\,, {Javlon Rayimbaev$^{3}$$^4$$^{5}$\thanks{javlon@astrin.uz}}\,, {Furkat Sarikulov$^{6}$\thanks{furqatsariquloff@gmail.com}}\,, and {Ozodbek Abdurakhmonov$^{7}$$^{8}$\thanks{ozodbekabdu@gmail.com}} 
	\vspace{0.2cm} \\
	$^1$College of Mathematics and Statistics, Shenzhen University, Shenzhen 518060, China\\
	$^2$ College of Physics and Optoelectronic Engineering, Shenzhen University, Shenzhen 518060, China\\
	$^{3}$School of Mathematics and Natural Sciences, New Uzbekistan University, Movarounnahr Str. 1,\\ Tashkent 100000, Uzbekistan \\ 
	$^{4}$University of Tashkent for Applied Sciences, Gavhar Str. 1, Tashkent 100149, Uzbekistan \\
	$^{5}$National University of Uzbekistan, Tashkent, Uzbekistan \\
	$^{6}$Ulugh Beg Astronomical Institute, Astronomy Str. 33, Tashkent 100052, Uzbekistan\\
	$^{7}$nstitute of Fundamental and Applied Research, National Research University TIIAME,\\ Kori Niyoziy 39, Tashkent 100000, Uzbekistan  \\
	$^{8}$PDP University, Yangi Sergeli Str. 12, Tashkent 100022, Uzbekistan
		}
	\date{}
\maketitle
\begin{abstract}
One of the most important and actual issues in relativistic astrophysics is testing gravity theories and obtaining constraint values for the parameters of black holes using observational data. In this research, we aimed to explore the optical features of a Kerr--Newman black hole model in the presence of a quintessential field, which may be a candidate for a dark-energy model with a nonzero cosmological constant. First, we obtain the equations of motion for photons using the Hamilton--Jacobi formalism. We also study the horizons and shapes of the apparent regions of the photon region around the said black hole. In various scenarios, we investigate shadows cast by the black hole using celestial coordinates. Furthermore, we analyze the effects of the quintessential field and black hole charge on the shadow radius and distortion. Furthermore, we look into the constraints on the spin and charge of supermassive black holes M87$^*$ and Sagittarius A$^*$ for different values of the quintessential field using their shadow size measured by the Event Horizon Telescope Collaboration. Finally, we study the effects of the quintessential field, black hole spin and charge on its energy emission rate by Hawking radiation and compare our results with those of the available literature.
\end{abstract}

\section{Introduction}\label{introduction}
According to recent cosmological discoveries, the centers of nearly all
galaxies could host supermassive black holes (BHs). Astrophysical evidence, in
particular, strongly implies the existence of a BH Sagittarius (Sgr)
A$^{*}$ at the center of our galaxy~\cite{Haroon:2018ryd}. 
In 2019, using very long baseline interferometry (VLBI), the Event Horizon
Telescope (EHT) project discovered the first shadow picture of a supermassive
BH in the center of galaxy
M87$^{*}$~\cite{EHT:2019dse,EHT:2019uob,EHT:2019jan,EHT:2019ths}. 
The second picture recently shows how an astrophysical environment, such as the
magnetic field, alters the BH image~\cite{EHT:2021bee}. Despite other evidence,
this is the most decisive evidence for the existence of gigantic BHs, leading
to new paths for BH research.

The discoveries mentioned above, particularly the EHT project's visualization
of the M87{$^{*}$}'s silhouette~\cite{EHT:2019dse,2019PhRvD.100d4057B},
caught the interest of many researchers to investigate BH
shadows~\cite{2019PhRvD.100b4020V}. Henceforth, this is a significant area of
research to test general relativity (GR) and modified theories of gravity.
Its mathematical
aspects~\cite{Synge:1966okc,Grenzebach:2014fha,Abdujabbarov:2015rqa,2016PhRvD..94b4054A,Stuchlik:2018qyz,Ovgun:2018tua,Dey:2020haf,Zahid2021ChJPh72,2022ChJPh..80..148M,2022EPJC...82..831A,2023PDU....4101248D,Zahid2023EPJC83,2023ChPhC..47b5102A,2024ChPhC..48a5103Y,2023EPJC...83..995A}
established during the years may be investigated using observable data, which
can assist us in polishing the mathematical models of these remarkable
astrophysical objects. The discovery validates previous mathematical
investigations supporting the presence of BHs in our universe and gives a
feasible means to check one of GR's core predictions. 
From ongoing data collection using the Thirty-Meter
Telescope~\cite{sanders2013thirty}, the Next Generation Very Large
Array~\cite{hughes2015next}, and the Black Hole Cam
project~\cite{Goddi:2016qax}, we anticipate an ideal chance during the upcoming
years to gain more insight into the strong gravity regime. To have information
on BHs, the study of particle dynamics and the approaches of null geodesics has
been an engaging problem in astrophysics~\cite{Khan2021ChJPh70,khan2023EPJC83}.
{Numerous scholars have looked at the geodesic structure of a BH
spacetime~\cite{Stuchlik:1999qk,Tursunov-KerrPRD,Khan2024EPJC291J,Rayimbaev:2020hjs,2021EPJC...81..419M,Khan2021AIPC,Khan2024EPJC84,2024ChJPh..88...55J},
as well as compact stars and wormholes~\cite{2021PhyS...96j5008M,2024ChJPh..88..938M,2024ChJPh..88...32M,2024ChJPh..87..751K},
as they are of considerable curiosity and have the potential to provide
important information while also revealing the effective layout of background geometry.}

The photon sphere is a spherical surface that revolves around a BH and contains
all conceivable circular orbits of a photon. However, rotating BHs do not have
a sphere; rather, they are an area separated from the photon sphere. It
describes an area occupied by light-like spherical geodesics. The
asymptotically twisted rays of light in one of those light-like spherical
geodesics correspond to the outermost edge of the
shadow~\cite{2020arXiv200903778C}. As a result, the features of the photon
regime are critical for studying BH shadows.  Bardeen proposed the concept of
BHs shadow~\cite{Bardeen:1973tla} after deducing that a spacetime BH had
$r_{shadow}=5.2M$ (shadow radius) on the background's light source. Subsequently, many
researchers observed that the magnitude of various spinning BH shadows is
almost identical to that found
earlier~\cite{Takahashi:2005hy,Hioki:2009na,Khan2020PDU}.
BH seems like a black disk known as a BH silhouette for an outside observer.
The shadows of the nonspinning BHs appeared to be a perfect circular disk,
whereas the shadows of the revolving BHs deformed in the direction of BH's
rotational axis~\cite{Bardeen:1973tla,Perlick:2018iye}. {Recently, Mustafa
et al.~\cite{2022ChPhC..46l5107M} examined the shadows and weak gravitational 
lensing by Schwarzschild BH under the influence of the quintessential field.}

In the last few decades, analyzing our universe's invisible aspects has piqued
numerous researchers' curiosity. Several cosmological datasets revealed that
dark energy (DE) is the critical instrument of the accelerating expansion of
our universe~\cite{Peebles:2002gy}. Around 68\% of our observable universe is
compiled of DE~\cite{Caldwell:2009zzb}, which might be described by a repulsive
cosmological constant $\Lambda>0$ or by the quintessence
fields~\cite{Ostriker:1995su,Caldwell:1997,Faraoni:2000wk,Adami:2012uv,2022ApJ...941..170M}.
The parameter $\Lambda$ could be considered homogeneous, with the same value
of ($\Lambda\approx 1.3\times 10^{-56} \text{ cm}^{-2}$) in the space~\cite{Stuchlik:2005euw}. In addition to the
cosmological constant, quintessence is another key component of
DE~\cite{copeland2006dynamics}. It is a dynamical and in-homogeneous scalar
field with negative pressure, characterized by the expression $\omega = p/\rho$,
known as the equation of state, in which $p$ and $\rho$,
respectively, indicate the pressure and energy density of the field. Based on
state parameter $\omega$, there exit three cases: corresponding to the
phantom energy ($\omega < -1$), the cosmological constant ($\omega = -1$), and the
quintessence $\omega \in (-1, -\frac{1}{3})$~\cite{1999PhRvD..59l3504S}. The parameter of state is
the most essential characteristic for understanding the concept of DE. As a
result, knowing the equation of state is critical for determining
DE~\cite{2018ApJ...857....9D}.
The components of DE, namely, the cosmological parameter and/or the
quintessential fields of BHs, significantly impact spacetime geometry. As a
result, an asymptotic geometry of the spacetime BH takes the form of an
asymptotic de-Sitter BH in the existence of a cosmological
constant~\cite{Stuchlik:1999qk,Xu:2016jod}.
Consequently, DE or dark matter must be considered in the solutions of BHs.

In recent years, BH, enveloped by the quintessential field, has attracted
numerous numbers of scientists. For example, Kiselev~\cite{Kiselev:2002dx}
postulated the Schwarzschild BH enclosed by a quintessential field, which was
later expanded by Toshmatov et al.~\cite{Toshmatov:2015npp} to the Kerr-like BH;
the quasinormal modes, thermodynamics, and phase transition around the Bardeen
BH enclosed by quintessence were studied in~\cite{Saleh:2017vui}. Previously,
by studying BH shadow in the presence of a cosmological constant and
quintessential field, we observed that $\gamma$ and $\Lambda$ contribute
to the shadow radius, whereas diminishes the distortion effect at higher
values~\cite{Khan2020PDU,2022ChJPh.78.141K}.
In addition, various authors have hypothesized BHs surrounded by quintessence
with a non-vanishing cosmological
constant~\cite{Azreg-Ainou:2014lua,Hong:2019yiz}.

The key objective of this article is to study the optical features of a Kerr--Newman (KN) BH in the quintessential field with a nonzero cosmological constant (KNdSQ BH). In the following Section~\ref{sec:02}, we will briefly review the spacetime geometry of a KNdSQ BH. Section~\ref{sec:03} explores the characteristics of the photon region around a KNdSQ BH. The fourth  Section~\ref{sec:04} of our article is devoted to investigating BH shadow and observables. Section~\ref{sec:Constraints} of our article deals with the constraints from EHT observations. We investigate the energy emission rates in Section~\ref{sec:eemission}. In the last Section~\ref{sec:conclusion}, we conclude our study with closing remarks.
\section{Spacetime metric of the KNdSQ BH}\label{sec:02}
{The KN BH provides a more realistic representation of astrophysical BH as
	most of them, such as spinning and magnetic BHs, are expected to have both
	angular momentum and charge~\cite{stephani2009exact}. It gives an in-depth
	understanding of the various characteristics BHs may exhibit in the cosmos. The
	KN BH is a generalization of the Kerr metric, which is extended to include
	charge, enabling a broader exploration of the BH characteristics and their
	implications for observational studies. Moreover, considering the
	quintessential field, it becomes more interesting and provides a broader
	examination of BH features.
	Thus, in the present section, we are interested in investigating the KNdSQ BH,
	which is the solution of the Einstein--Maxwell equations and has the form in
	Boyer--Lindquist coordinates~\cite{Xu:2016jod}}
\begin{eqnarray}\label{e1}
	ds^2=\frac{\Delta _\theta \sin^2\theta}{\rho^2}{\left(a\frac{dt}{\Sigma}-\left(a^2+r^2\right)\frac{{d\phi}}{\Sigma }\right)^2}+\frac{\rho^2}{\Delta_r}{dr}^2 +\frac{\rho^2}{\Delta_\theta}{d\theta}^2  - \frac{\Delta _r}{\rho ^2}{ \left(\frac{dt}{\Sigma}-a\sin^2\theta \frac{d\phi}{\Sigma }\right)^2},
\end{eqnarray}
with
\begin{eqnarray}
	\nonumber
	\Delta_r &= &r^2-2Mr+a^2+Q^2-\gamma r^{1-3\omega}-\frac{\Lambda}{3}(r^2+a^2)r^2, \\\nonumber
	\rho^2 &=&r^2+a^2 \cos^{2}\theta, \quad \Delta_\theta = 1+ \frac{a^2}{3}\Lambda \cos^2\theta, \quad \Sigma = 1+\frac{a^2}{3}\Lambda.
\end{eqnarray}
Here $M$, $a$, and $Q$, respectively, stand for BH's mass, spin, and charge. The parameter $\Lambda$ represents the cosmological constant, whereas $\gamma$ defines BH's quintessence field's 
intensity. Moreover, $\omega$ describes the equation of state given by $p = \omega \rho$, where $\rho$ and $p$ respectively represent the energy density and pressure of quintessence.
By inserting $\gamma=0$ and replacing $Q^2$ by $q$, the metric
in Eq.~\eqref{e1} reduces to the braneworld BH with a nonzero cosmological
constant~\cite{2022ChJPh.78.141K}; simplifies to the KN case by substituting
$\gamma=\Lambda=0$; reduces to the standard Kerr spacetime with $\gamma=\Lambda=Q=0$; to the RN
BH with $\gamma=\Lambda=a=0$; and finally to the Schwarzschild BH by setting
$\gamma=\Lambda=a=Q=0$.

Horizons of the above metric \eqref{e1} are the null hypersurfaces induced by the BH's null geodesics, which can be recovered from the solution of  $\Delta_r=0$, as
\begin{equation}
	\label{EH}
	r^2-2Mr+a^2+Q^2-\gamma r^{1-3\omega}-\frac{\Lambda}{3}(r^2+a^2)r^2=0.
\end{equation}
Xu and Wang previously detailed how the parameters $\gamma$, $a$,
$Q$, $\Lambda$, and $\omega$ define the number of
horizons~\cite{Xu:2016jod}. Furthermore, one should note that its perspectives
differ from those of the standard Kerr BH.  

By setting $g_{tt} = 0$, the ergosphere may be determined, or equivalent by solving
\begin{equation}
	\label{SLS}
	a^2\sin^2{\theta \Delta_{\theta}}-\Delta_r=0,
\end{equation}
it defines the static limit surface (SLS), which is also affected by $\theta$. On letting $\gamma=\Lambda=Q=0$, the SLS simplifies to the case of standard Kerr BH and accepts the solution
\begin{equation}
	r_{SLS\pm}=M\pm \sqrt{M^2-a^2\cos^2\theta}.
\end{equation}
\section{Photon region}\label{sec:03}
This section explores the photon region's characteristics around a KNdSQ BH, with the condition of $p^{\mu}p_{\mu}=-m^2$. Henceforth, by applying the technique of separation of variables, the corresponding geodesics motion of metric \eqref{e1} could be expressed by the Hamilton--Jacobi equation as
\begin{figure*}[ht!]
	\begin{minipage}[b]{\textwidth}\hspace{-0.1cm} 
		\includegraphics[width=0.32\textwidth]{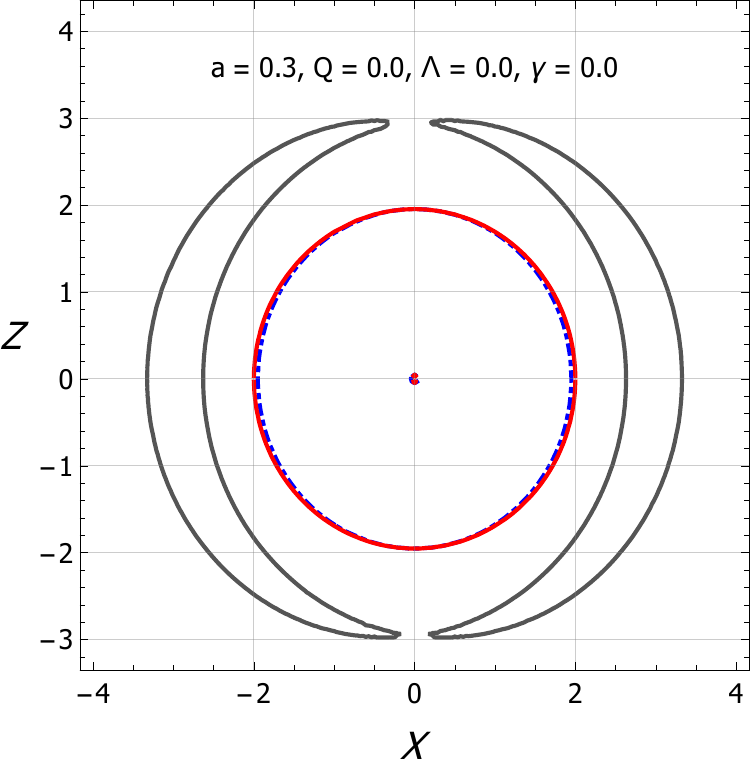}
		\includegraphics[width=0.32\textwidth]{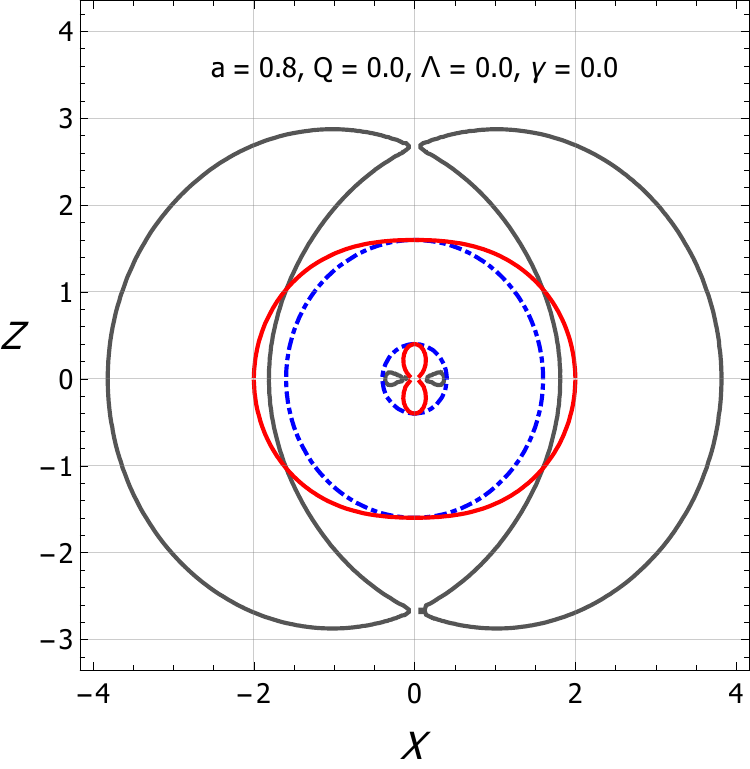}
		\includegraphics[width=0.32\textwidth]{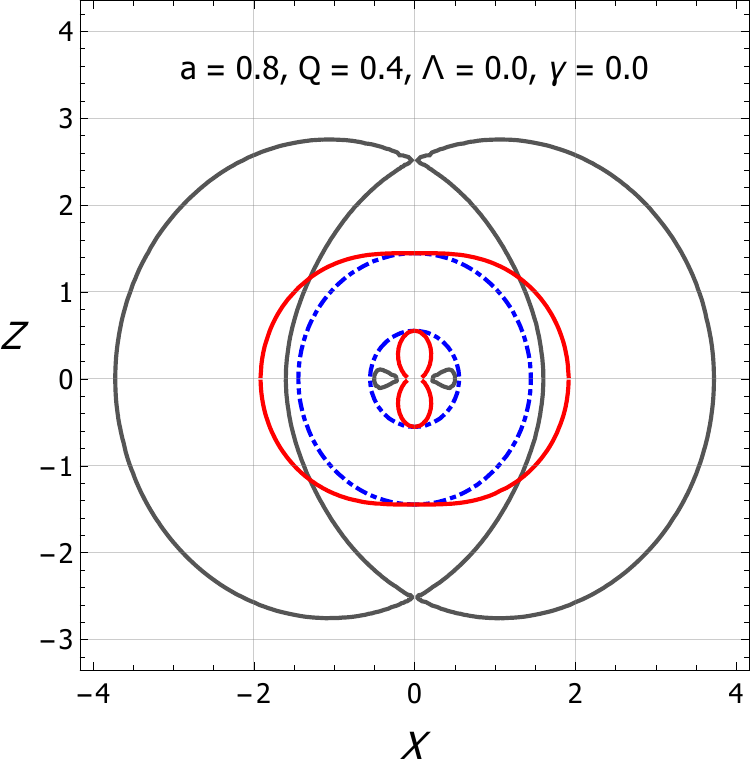}
	\end{minipage}
	\begin{minipage}[b]{\textwidth} 
		\includegraphics[width=0.32\textwidth]{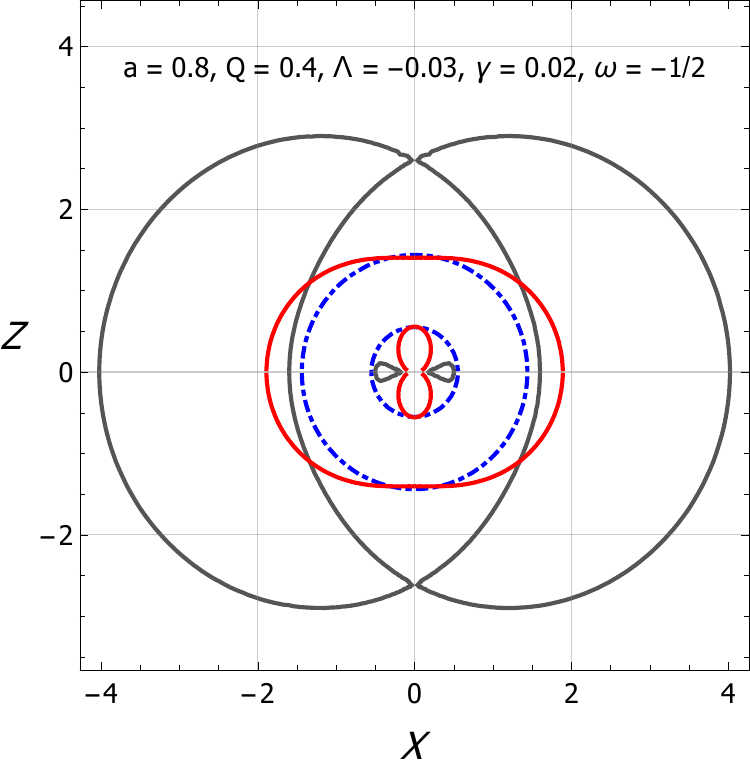}
		\includegraphics[width=0.32\textwidth]{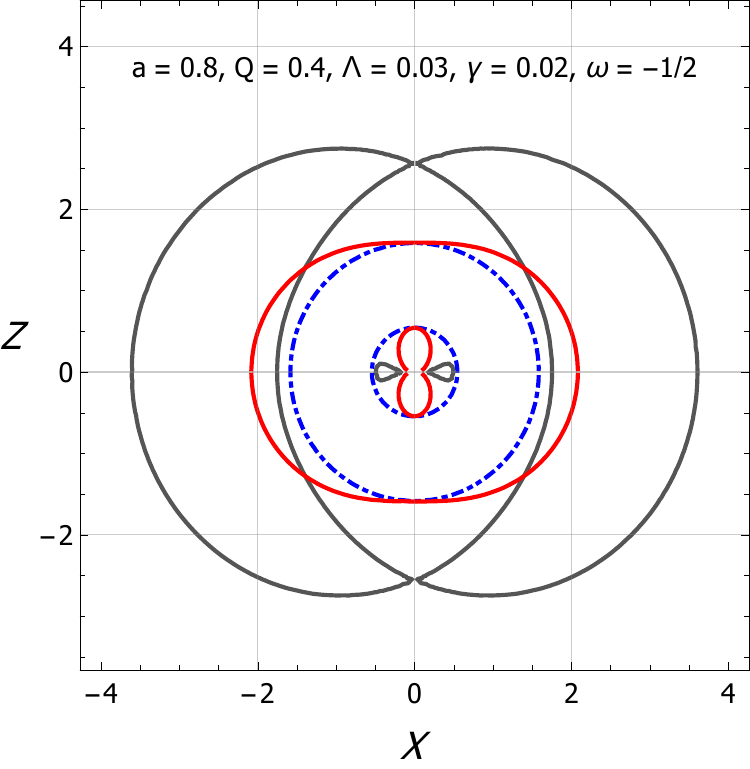}
		\includegraphics[width=0.32\textwidth]{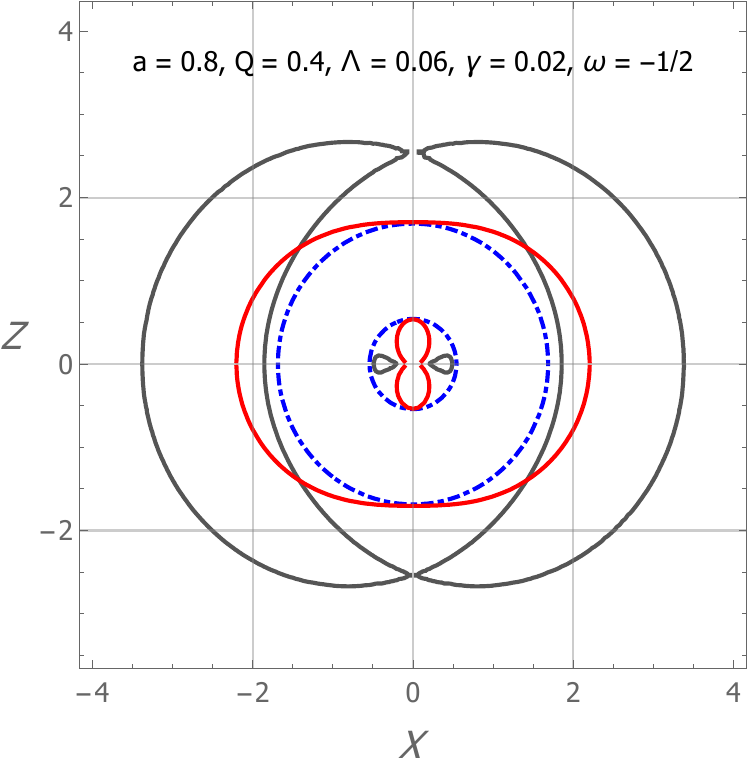}
	\end{minipage}
	\begin{minipage}[b]{\textwidth}
		\includegraphics[width=0.32\textwidth]{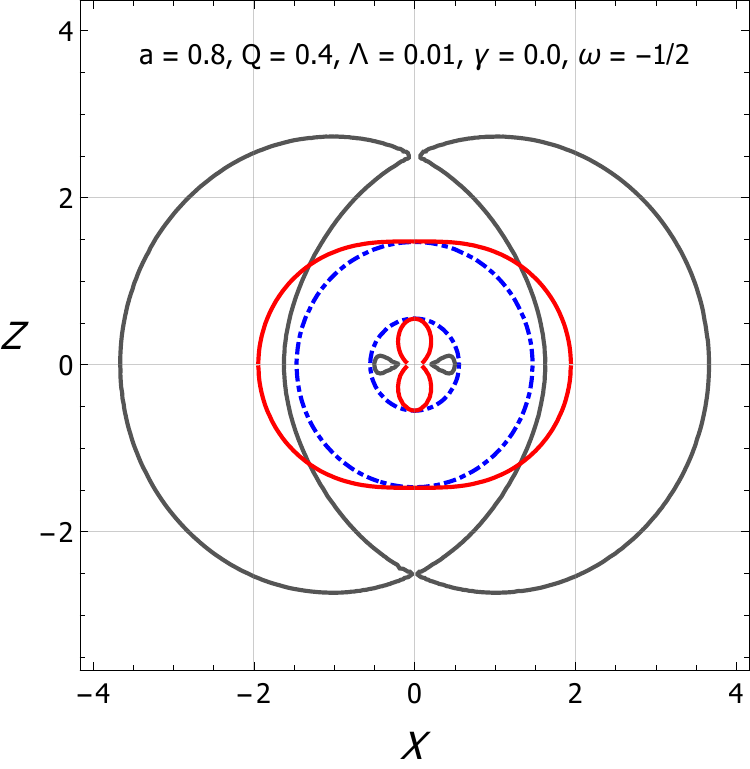}
		\includegraphics[width=0.32\textwidth]{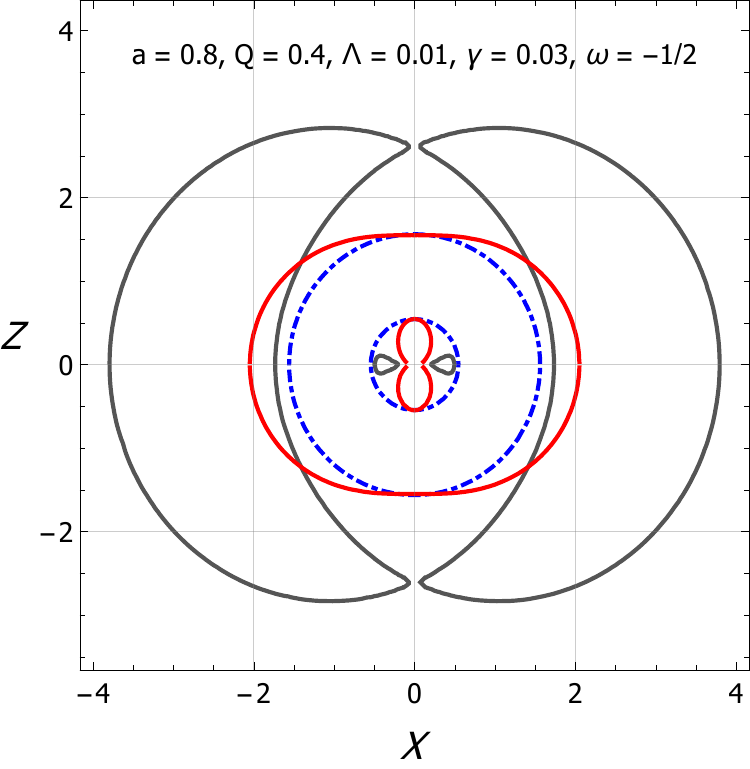}
		\includegraphics[width=0.32\textwidth]{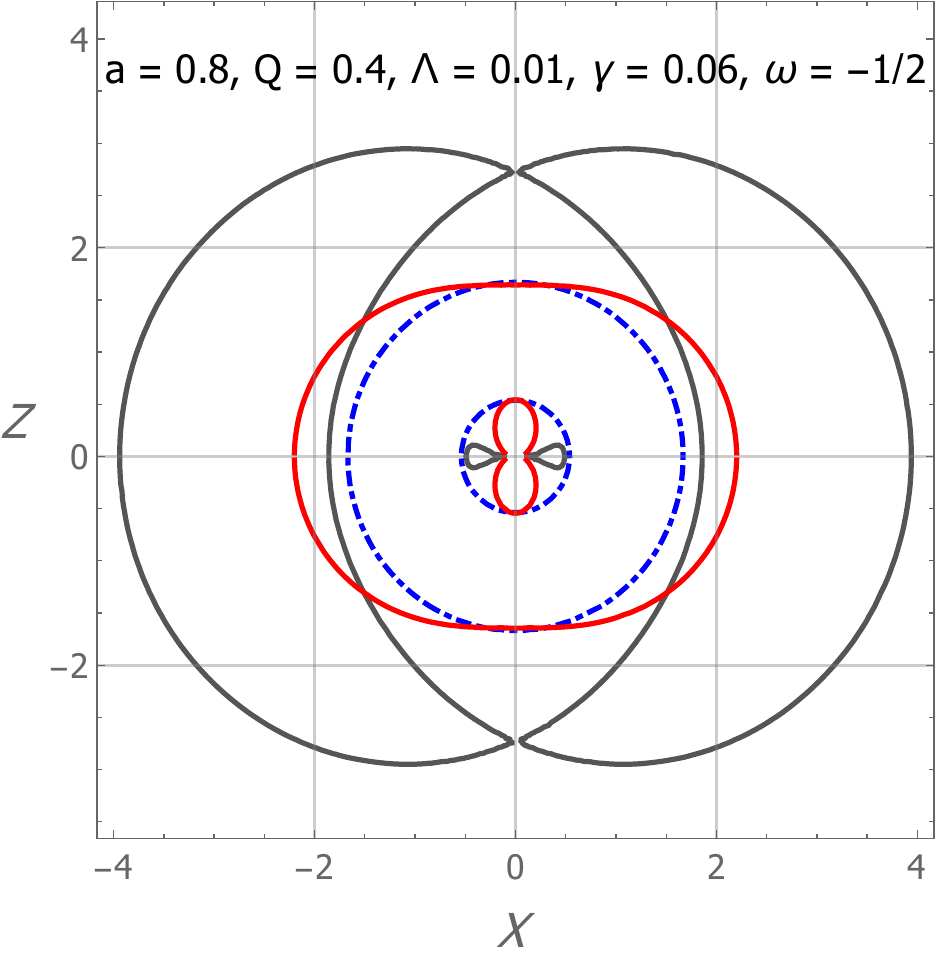}
	\end{minipage}
\caption{The apparent shapes of the photon region (area between the gray cures on both sides of the origin), Horizons (the inner and outer blue dashed curves, respectively define the Cauchy and event horizon, whereas the inner and outer red curves, respectively represent the inner and outer subsurfaces). Moreover, the area between the event horizon and outer ergosurface is called the ergoregion.}\label{Pregion1}
	\end{figure*}
\begin{equation}
	\label{BS1}
	\mathcal{H} = -\frac{\partial S}{\partial \tau}= \frac{1}{2}g_{\mu\nu} \frac{\partial S}{\partial{x}^{\mu}} \frac{\partial S}{\partial{x}^{\nu}}=-\frac{1}{2}m^2,
\end{equation}
in Eq.~\eqref{BS1}, $\mathcal{H}$ and $S$, respectively, denote the Hamiltonian and Jacobi action, while $m^2$ is the rest mass of the particles. But when considering photon orbits, we substitute $m^2=m_0^{2}=0$. In light of the four-momentum $p_\mu={\partial S}/{\partial{x}^{\mu}}$, and by making use of Carter's
separability prescription~\cite{Carter:1968rr}, the action could be separated
as
\begin{equation}
	\label{BS2}
	S=\frac{1}{2}m_0^2\tau-\mathcal{E}t+L_z\phi+S_r(r)+S_\theta(\theta).
\end{equation}
The symmetry of spacetime allows us to define the corresponding conserved energy and angular momentum as follows:
\begin{eqnarray}
	\label{BS8}
	\mathcal{E}&=&-p_{t}=g_{tt}\dot{t}+g_{t\phi} \dot{\phi},\\ \label{BS9}
\mathcal{L}&=&p_{\phi}=g_{t\phi}\dot{t}+g_{\phi\phi}\dot{\phi}.
\end{eqnarray}
On considering the geometry of the KNdSQ BH, the equations of motion in terms
of the four differential equations in the $r-\theta$ plane can be expressed
as~\cite{2018EPJC...78}
\begin{eqnarray}\label{BS10}
	\rho^2\frac{dt}{d\tau}&=& \frac{a}{\Delta_\theta}(\Sigma \mathcal{L}-a \mathcal{E}\sin^2\theta) +\frac{(a^2+r^2)}{\Delta_r} (\left(a^2+r^2)\mathcal{E} -a\Sigma \mathcal{L} \right),\\\label{BS11}
	\rho^2\frac{dr}{d \tau}&=&\sigma_r \sqrt{\mathcal{R}(r)},\\\label{BS12}
	\rho^2\frac{d \theta}{d \tau}&=& \sigma_\theta \sqrt{\Theta(\theta)},\\\label{BS13}
	\rho^2\frac{d\phi}{d\tau}&=&\frac{\Sigma}{\Delta_\theta} \left (\Sigma\mathcal{L}\csc^2\theta-a \mathcal{E} \right)+\frac{a\Sigma}{\Delta_r} (\left (a^2+r^2)\mathcal{E}
	-a\Sigma \mathcal{L}\label{BS14} \right).
\end{eqnarray}
In which $\sigma_r=\sigma_\theta=\pm$, with
\begin{eqnarray}
	\label{BS5}
	\mathcal{R}(r)&=&\left((a^2+r^2)\mathcal{E}-a \Sigma \mathcal{L}\right)^2-\Delta_r \left((\mathcal{L}-a\mathcal{E})^2+\mathcal{O} \right),\\\label{BS6}
	\Theta(\theta)&=&\mathcal{O}\Delta_{\theta} - \left(\mathcal{L}^2 \csc^2\theta - a^2 \mathcal{E}^2 \right)\cos^2\theta.
\end{eqnarray}
In the aforementioned expressions, $\mathcal{O}=\mathcal{K}-(a\mathcal{E}-\mathcal{L})^2$ denotes Carter's constant with
$\mathcal{K}$ being the constant of motion. The trajectories of photons around a
KNdSQ BH is governed by the above four Eqs.~\eqref{BS10}-\eqref{BS13}. These trajectories depend mainly on the impact parameters
$\xi=\mathcal{L}/\mathcal{E} $ and $\eta=\mathcal{O}/\mathcal{E}^2$~\cite{Chandrasekhar1983}. In principle, photons
have three different types of trajectories called scattering, spherical, and
plunging orbits; while in terms of $\xi$ and $\eta$, the radial
Eq.~\eqref{BS5} takes the form
\begin{equation}
	\label{BS14a}
	\mathcal{R}(r)=\frac{1}{\mathcal{E}^2}\left[ \left((a^2+r^2)- a\Sigma \xi \right)^2 -\Delta_r\left((a-\xi)^2+\eta \right) \right].
\end{equation}
The effective potential on the photon can be straight-forwardly obtained from Eq.~\eqref{BS14a}. Thereafter, the critical and unstable circular orbits could be acquired using the maximum effective potential obeying the following constraints,
\begin{equation}
	\label{BS15}
	\mathcal{R}(r)=\frac{\partial \mathcal{R}(r)}{\partial r}\mid_{r=r_0}=0.
\end{equation}
Here $r_0$ is the radius of photons or unstable circular orbits. Making use of Eq.~\eqref{BS15}, the celestial coordinates $\xi$ and $\eta$ can be expressed as
\begin{eqnarray}
	\label{BS16}
	\xi&=&\frac{\left(a^2+r^2\right) {\Delta^\prime_r}-4 r {\Delta_r}}{a \Delta^\prime_r \Sigma}, \\\label{BS17}
	\eta&=&\frac{1}{(a {\Delta^{\prime}_r} \Sigma)^2}\left( r^2 \left(8 \Delta_r \left(2 a^2 \Sigma^2 + r \Delta^\prime_r\right)-r^2 {\Delta^\prime_r}^2-16 \Delta_r^2\right) -a^2 \Delta^\prime_r (1-\Sigma)\left(a^2 \Delta^\prime_r (1-\Sigma)+2 r^2 \Delta^\prime_r-8 r \Delta_r \right) \right). 
\end{eqnarray} 
{It should be noted that the impact parameters listed above are essential, as they determine the boundary of photon orbits.} The expression of the photon region in the $(r, \theta)$ plane can be obtained by substituting the values of $\xi$ and $\eta$ (Eqs.~\eqref{BS16} and \eqref{BS17}) into $\Theta \geq 0$,
\begin{equation}
	\label{Pregion}
	\eta \geq \frac{1}{\Delta_{\theta}} \left( \xi^2 \csc^2{\theta}-a^2 \right) \cos^2{\theta}.
\end{equation}

The graphical behavior of photon region in various cases is presented in Fig.~\ref{Pregion1}. The first row of Fig.~\ref{Pregion1} provides a comparison
between the slowly rotating Kerr (left panel), the fast-rotating Kerr (middle
panel), and the Kerr--Newman BH (right panel). Here, one can see that both spin
and charge parameters considerably contribute to the area of ergoregion, as
well as the photon region backing the previous
findings~\cite{Khan2019PDU,Fathi:2020agx}. Our observation summarizes that a
fast-rotating charged BH has a greater area of the ergoregion and photon region
than the chargeless and slowly rotating BHs. On the other hand, we observed
from the second row of Fig.~\ref{Pregion1} that the negative value of cosmological
constant $\Lambda$ increases the photon region, while $\Lambda > 0$ results in
diminishing it, which shows a good agreement with the results
of~\cite{Chen:2021wqh}. Moreover, one should note that $\Lambda > 0$ enlarges the
event horizon and outer ergosruface (SLS). For the variation of the
quintessence parameter $\gamma$, we observe that $\gamma$ contributes to
both the inner and outer boundaries of the photon region. However, the inner
boundary increases rapidly compared to the outer boundary (see the third row of
Fig.~\ref{Pregion1}). Therefore, $\gamma$ results in a shrinkage of the area of
the photon region while contributing to the radii of the event horizon and SLS. 
\section{BH Shadow}\label{sec:04}            
\begin{figure*}[ht!]\centering
\includegraphics[width=0.315\textwidth]{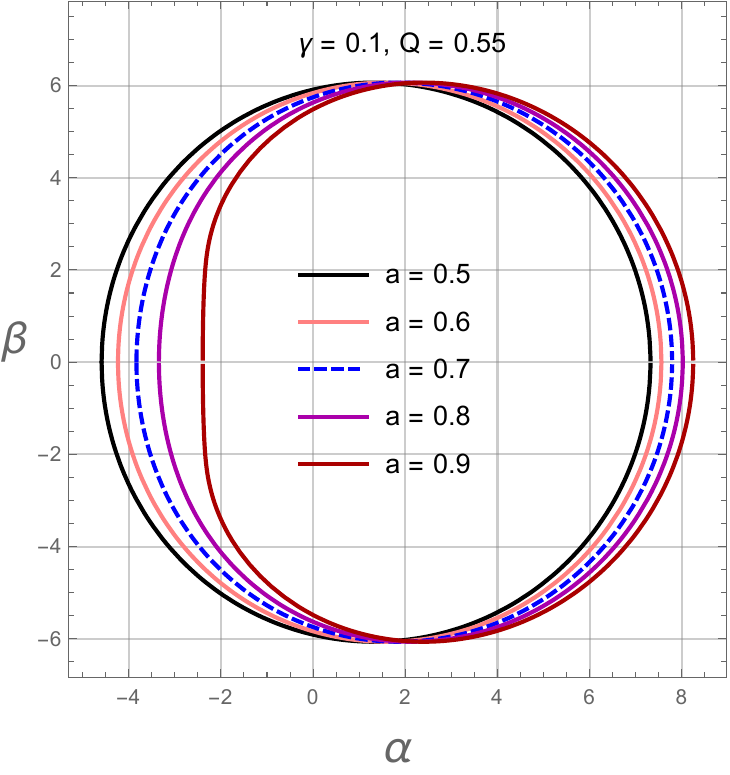}
\includegraphics[width=0.316\textwidth]{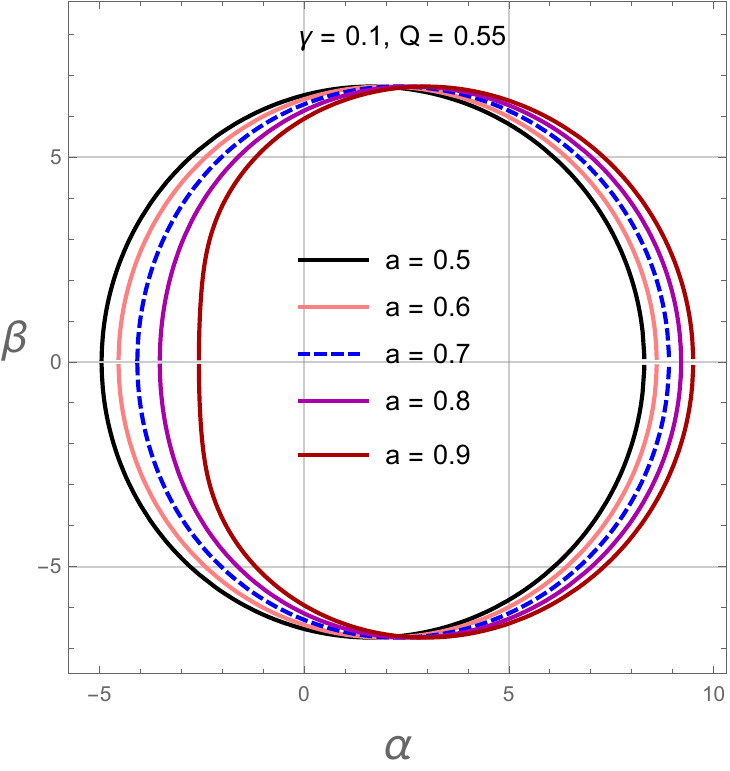}	\includegraphics[width=0.315\textwidth]{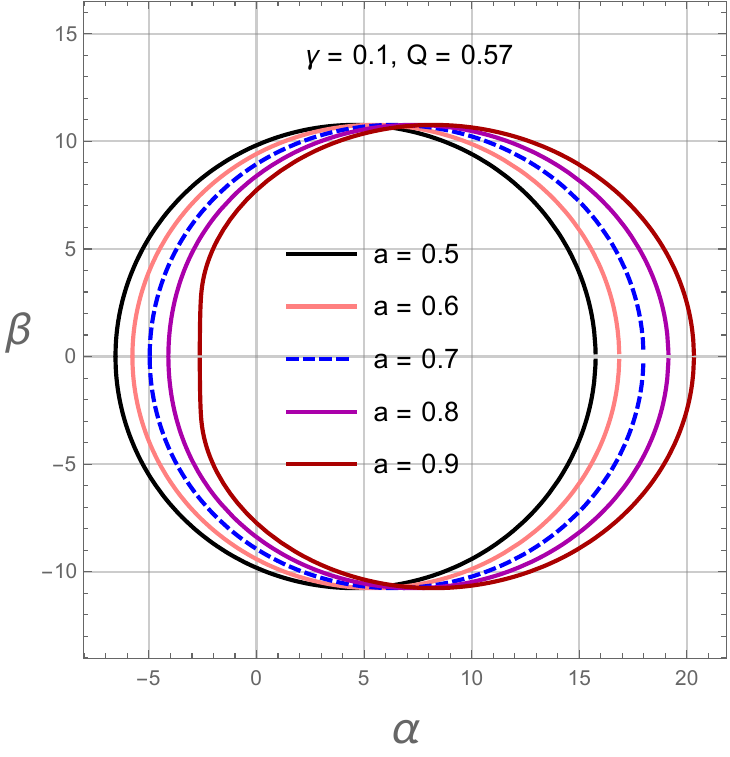}
\includegraphics[width=0.315\textwidth]{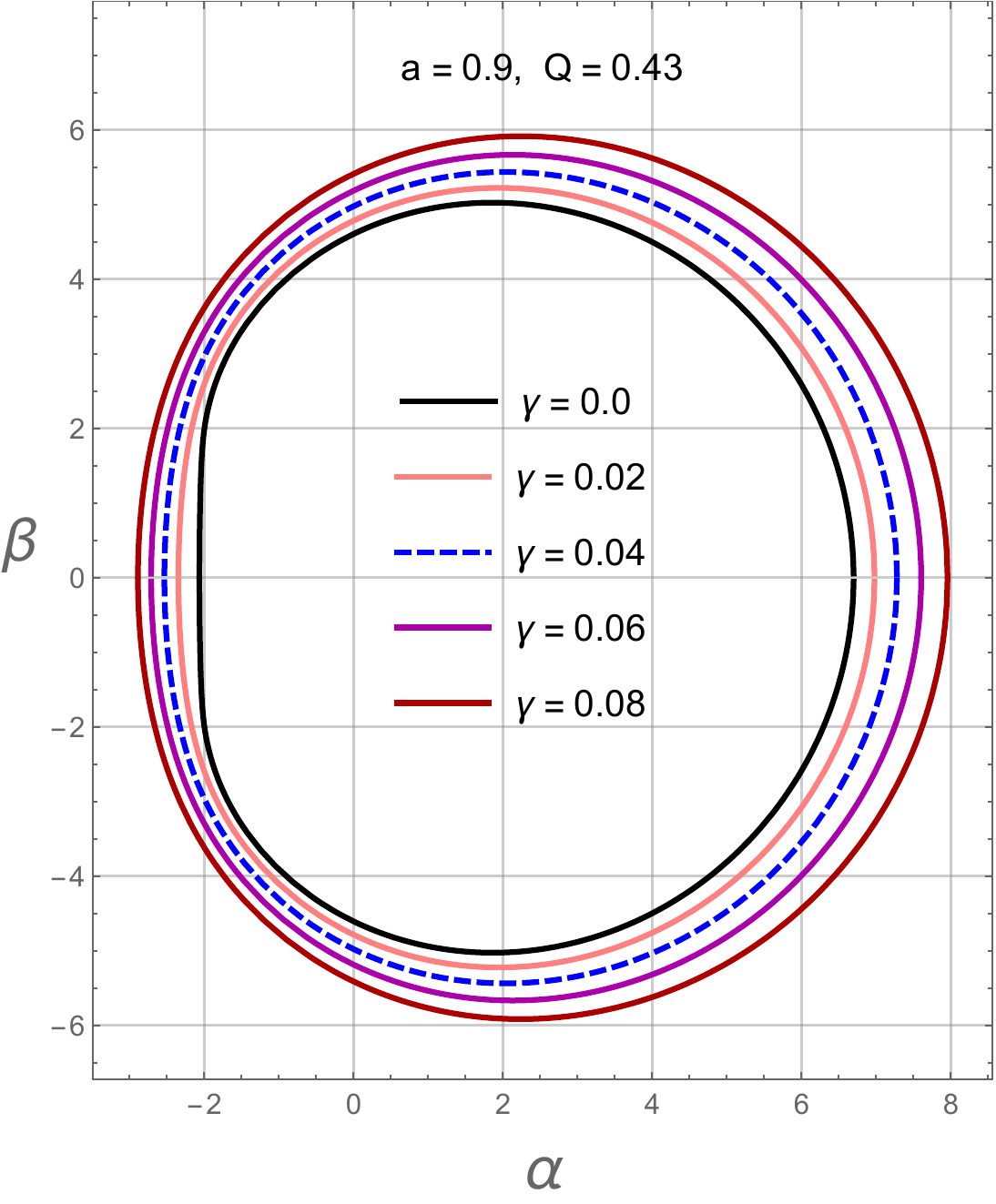}
\includegraphics[width=0.315\textwidth]{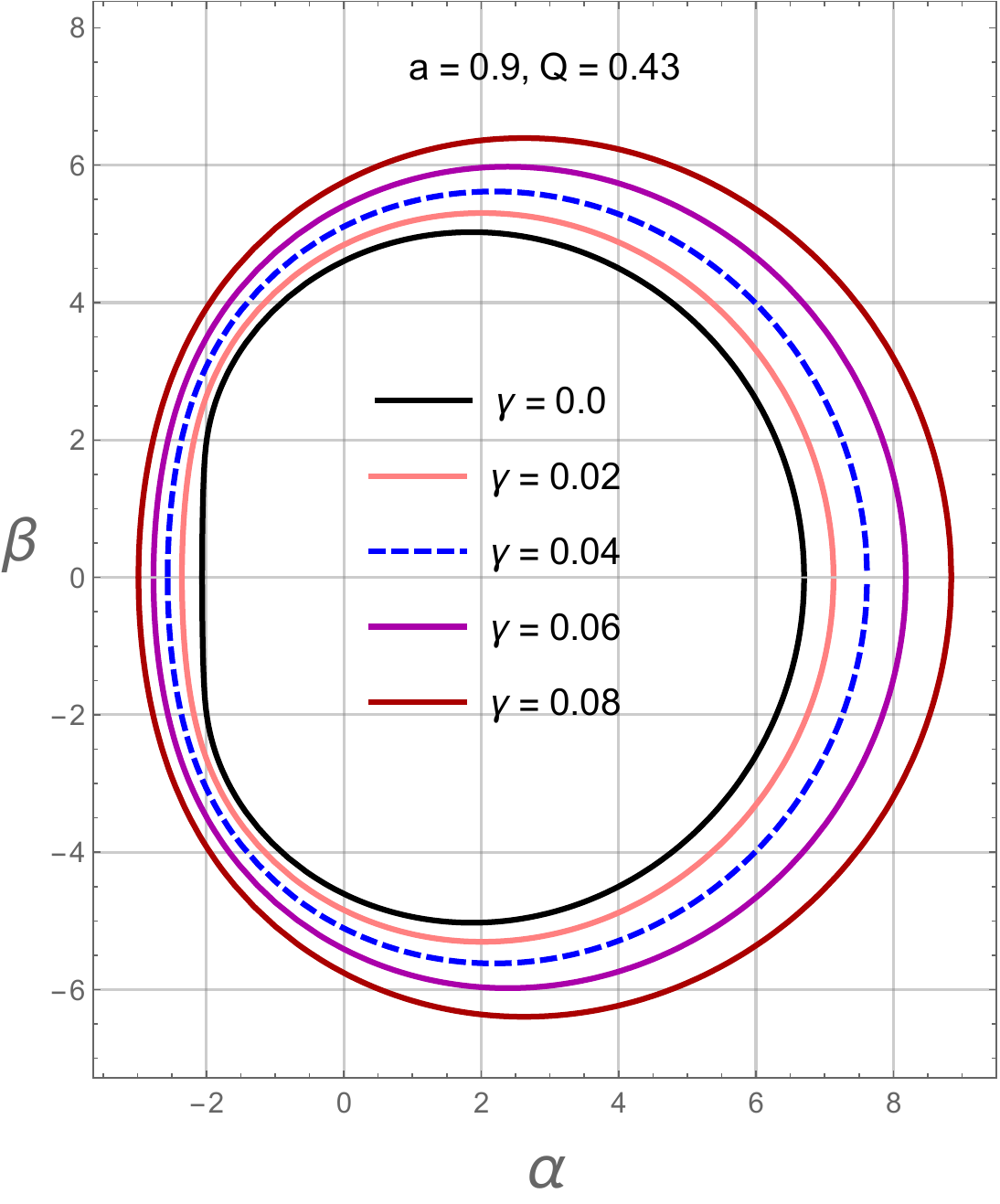}	\includegraphics[width=0.315\textwidth]{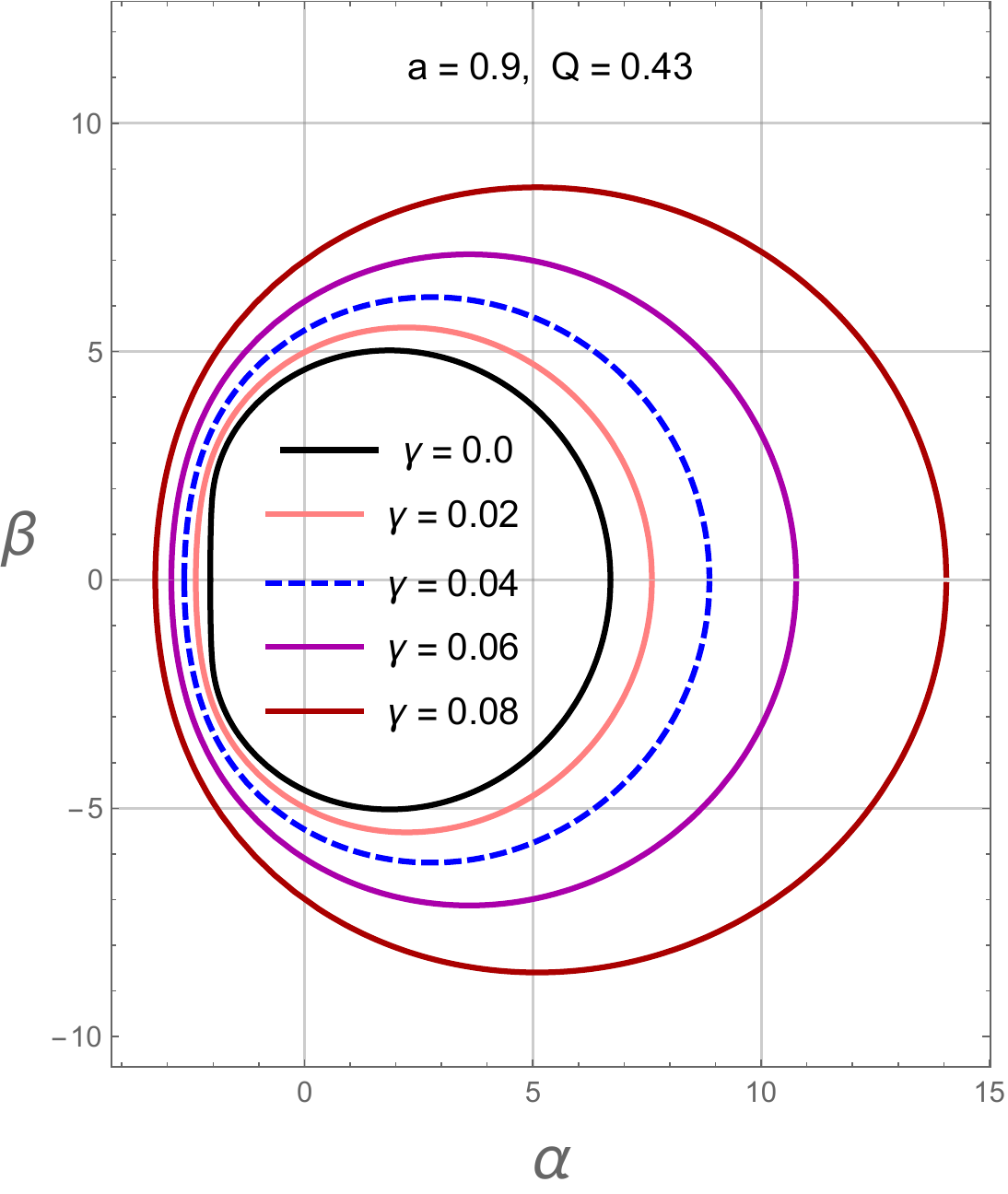}
\caption{Apparent shape of KNdSQ BH's shadow for the spin parameter (top row), and intensity of the quintessence field $\gamma$ (bottom row). The left, central and right panels are plotted for $\omega = -2/5$, $\omega = -1/2$ and $\omega = -2/3$, respectively.}\label{Shadow1}
\end{figure*}

This section is devoted to studying the shadow cast by the KNdSQ BHs. {The idea of BH shadow (a dark region in the sky)
was initiated by Bardeen in 1973~\cite{Bardeen:1973tla} for the Kerr spacetime. 
He deduced that BH casts a shadow of about $ R_{s} \simeq 5.2 M$ (representing the shadow radius) on a background light source. The boundary of the BH shadow indicates the threshold between escaped and captured photons; in other words, the spherical unstable photon orbits. In literature, such a boundary is known as the ``shadow'' or ``silhouette'' of a BH.}
In principle, the shadow of stationary BHs occurs just like a circular disk, whereas the spin parameter distorts the shadow of spinning BHs. The apparent
shape of a BH image can preferably be visualized with the help of celestial
coordinates $\alpha$ and $\beta$, which, under the assumption of
$\Lambda=0$, could be defined as~\cite{Hioki:2009na,Chandrasekhar1983}
\begin{eqnarray}\label{BS17a}
&&{\alpha}= \lim_{r_{\star} \to \infty} \left(-r_\star^2 \sin\theta \frac{d\phi}{dr} \right),\\
	\label{BS18}
&&{\beta}= \lim_{r_{\star} \to \infty}\left(r_\star^2 \frac{d\theta}{dr} \right).
\end{eqnarray}
\begin{figure*}[ht!]
\centering
\includegraphics[width=0.306\textwidth]{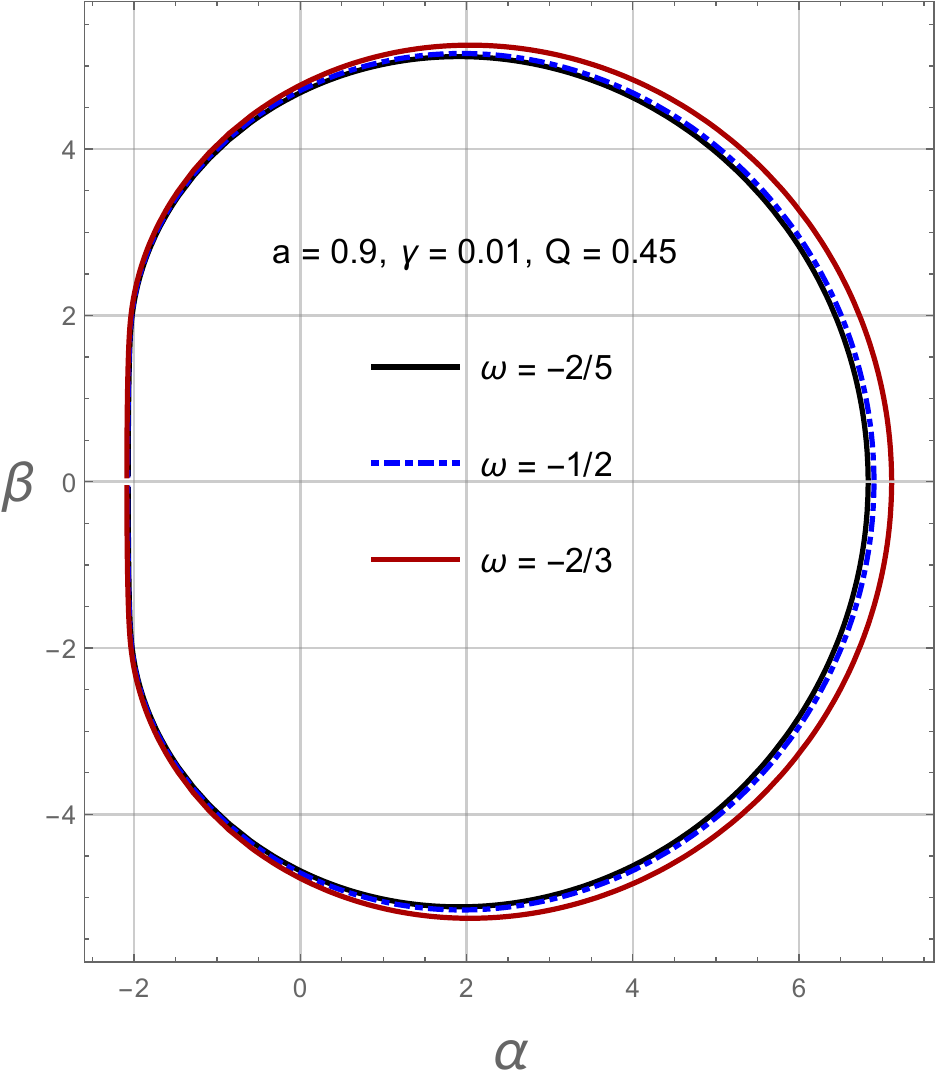}
\includegraphics[width=0.31\textwidth]{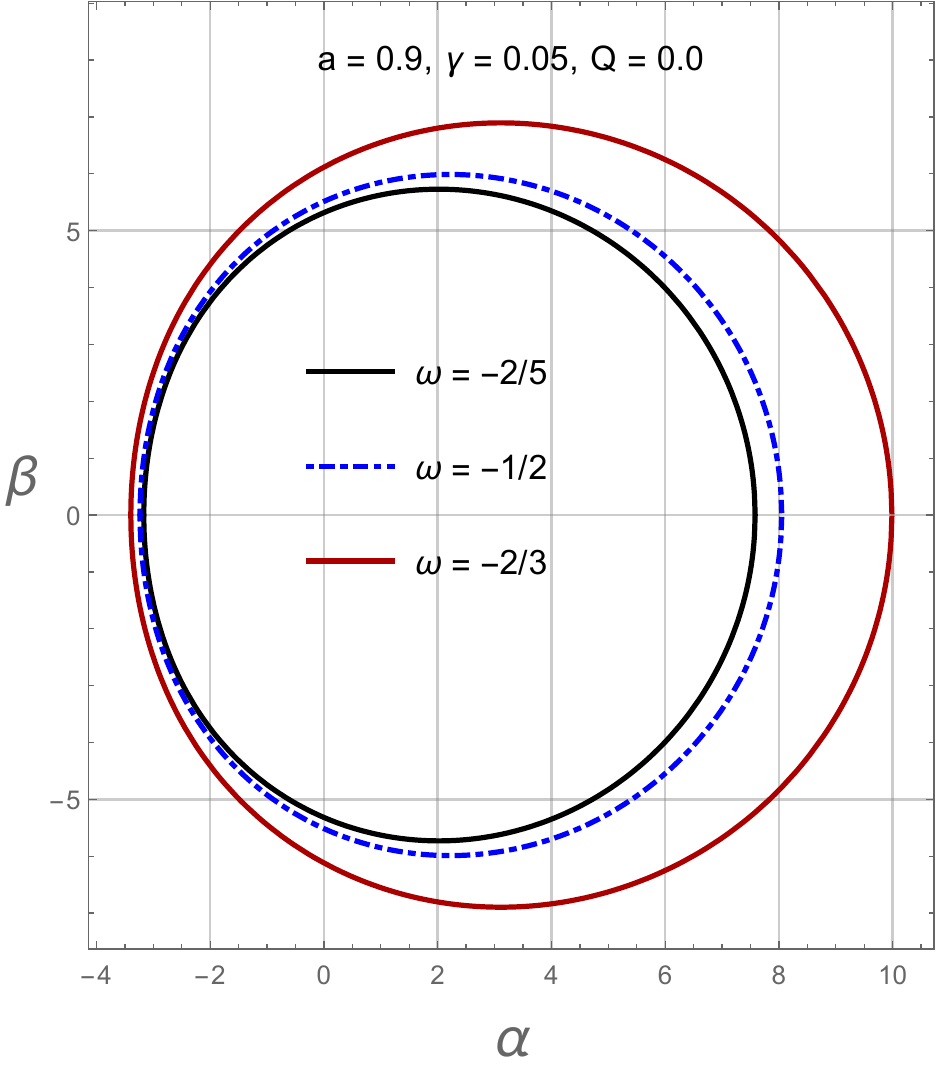}	\includegraphics[width=0.303\textwidth]{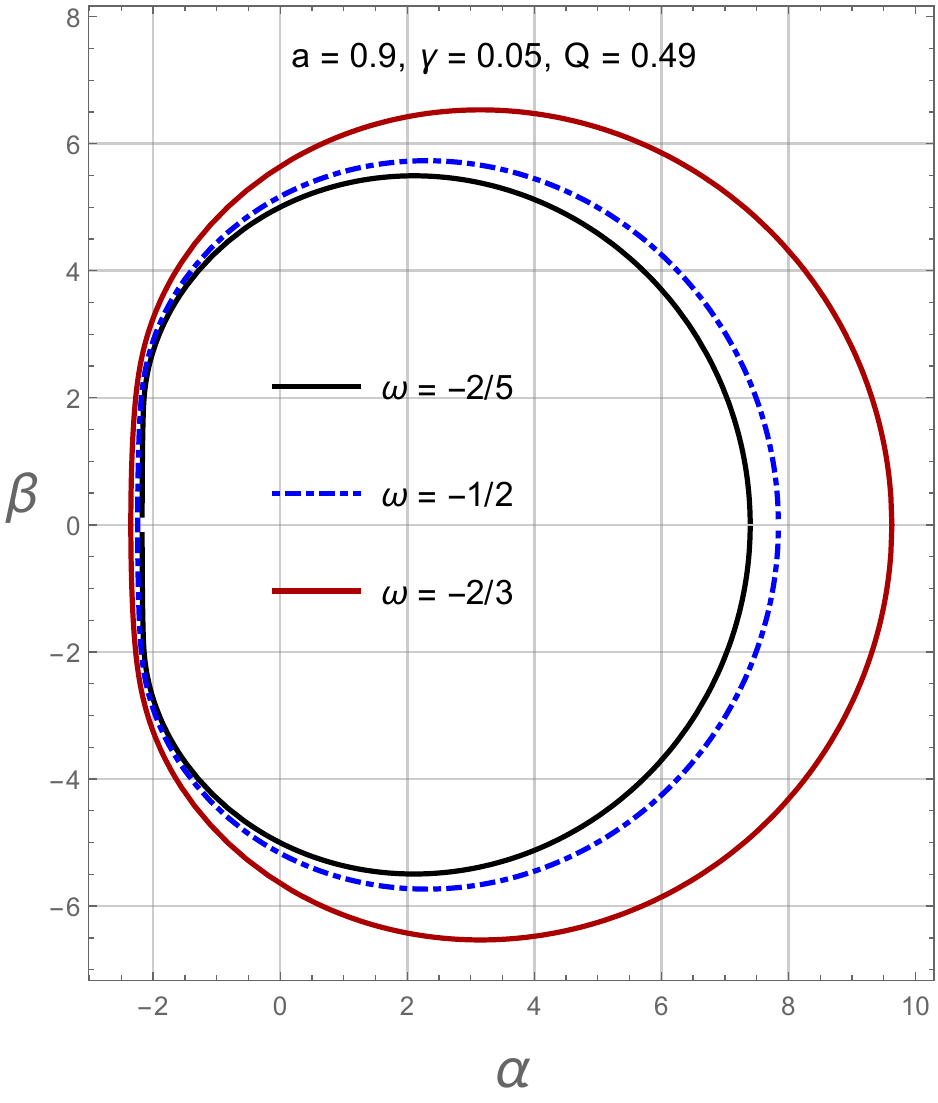}
\includegraphics[width=0.305\textwidth]{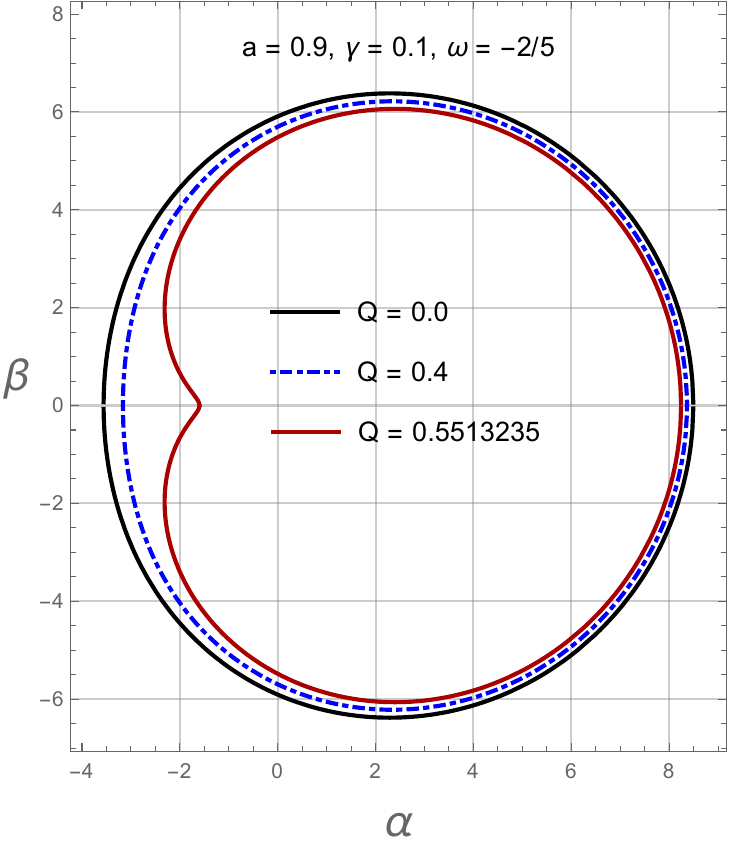}	\includegraphics[width=0.31\textwidth]{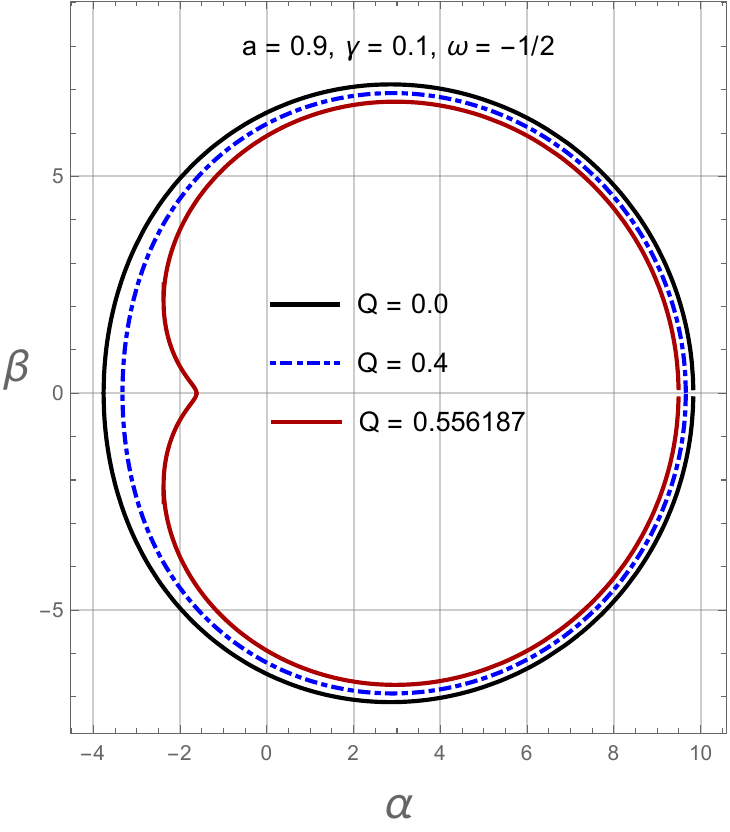}
	\includegraphics[width=0.32\textwidth]{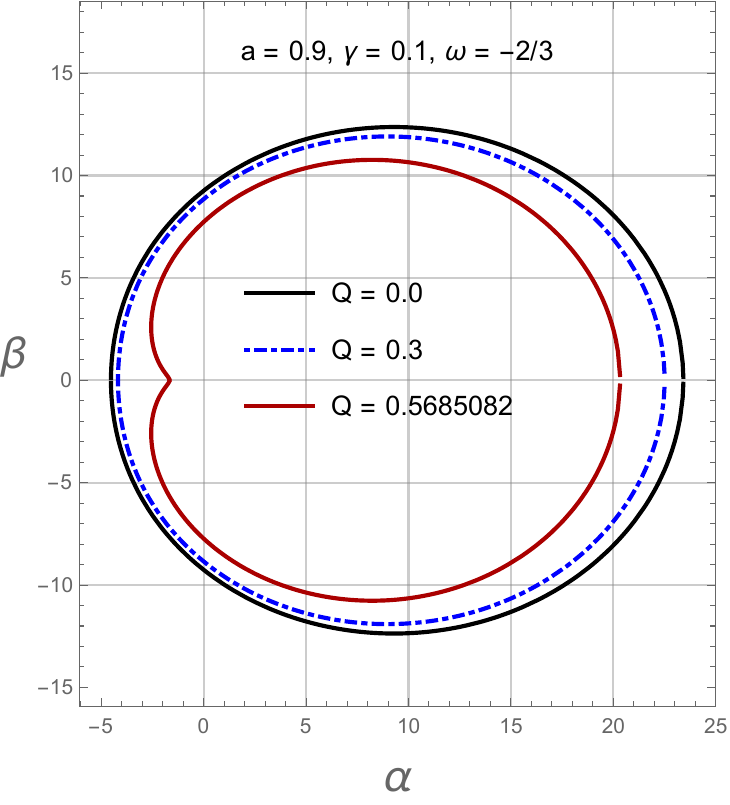}
\caption{Apparent shape of KNdSQ BH's shadow for the state parameter $\omega$ (top row) and BH charge (bottom row). }\label{Shadow2}
\end{figure*}

In the above expressions, $r_{\star}$ represents the distance between BH and the observer. At the same time, $ \theta$ denotes the angle of inclination between the observer's line of sight and the spinning axis of BH. By making use of the geodesic equations, the above Eqs.~\eqref{BS17a} and \eqref{BS18}, simplifying to
\begin{eqnarray}\label{BS19}
&&{\alpha}= -{\xi}{\csc \theta},\\
	\label{BS20}
&&{\beta}=\pm \sqrt{\eta+a^2\cos^2\theta-\xi^2 \cot^2\theta}.
\end{eqnarray}
Since BHs shadows are notable on the equatorial plane henceforth, with the substitution of $ \theta=\pi/2$, we obtain the following expressions:
\begin{eqnarray}\label{BS21}
&&{\alpha}=-\xi,\\
	\label{BS22}
&&{\beta}=\pm \sqrt{\eta}.
\end{eqnarray}   
Leaving $Q = \gamma=0$ the above expressions~\eqref{BS21} and \eqref{BS22}
yields
\begin{eqnarray}\label{BS23}
&&{\alpha}=\frac{a^2 (r+M)+r^2 (r-3 M)}{a (r-M)},\\
	\label{BS24}
&&{\beta}= \pm \frac{\sqrt{r^3 \left(4 a^2 M- r (r-3 M)^2\right)}}{a (r-M)}.
\end{eqnarray}
The aforementioned expressions,~\eqref{BS23} and \eqref{BS24}, are precisely the same as
those obtained by Hioki \& Maeda for the Kerr BH~\cite{Hioki:2009na}. 

The visual structure of a KNdSQ BH shadow is illustrated in Figs.~\ref{Shadow1} and \ref{Shadow2}. The apparent shape of the BH shadow illustrates that
due to the dragging effect, BH spin distorts as well as stretches the shadow it
casts in the direction of its axis of rotation, which aligns with the recent
findings~\cite{2024EPJC...84...63Y}. In addition, the BH charge contributes to
the distortion effect and distorts it considerably in the case of fast rotation
(see the first row of Fig.~\ref{Shadow1}). Moreover, the shadow of Kerr BHs
appeared wider than that of KN BHs.  The quintessence parameter $\gamma$
elongates the shadow radius while reducing the distortion effect, which becomes
negligible at higher values of $\gamma$ (see the second row of Fig.~\ref{Shadow1}). One should also note that $\gamma > 0$ keeps the shadow away
from distortion, and even a smaller value of $\gamma$ noticeably reduces
the distortion effect for details; see Fig.~\ref{Shadow1} second row. 

In addition to other parameters, the absolute value of $\omega$ also
elongates the BH shadow to the right. The last row of Fig.~\ref{Shadow2} shows
that the BH charge reduces the shadow radius and leads to a more distorted
shadow. At the same time, the distortion becomes very intensive at the critical
value of the BH charge (for details, see the second row of Fig.~\ref{Shadow2}),
which shows great consistency with the results of Ref.~\cite{2018EPJC...78}. On
summarizing the effects of DE energy on BH shadow, we can say that DE expands
the apparent shape of BH shadow, while its wideness also depends on the value
of $\omega$.

\subsection{Observables}\label{sec:5.1}
\begin{figure*}[ht!]\centering
	\includegraphics[width=0.5\textwidth]{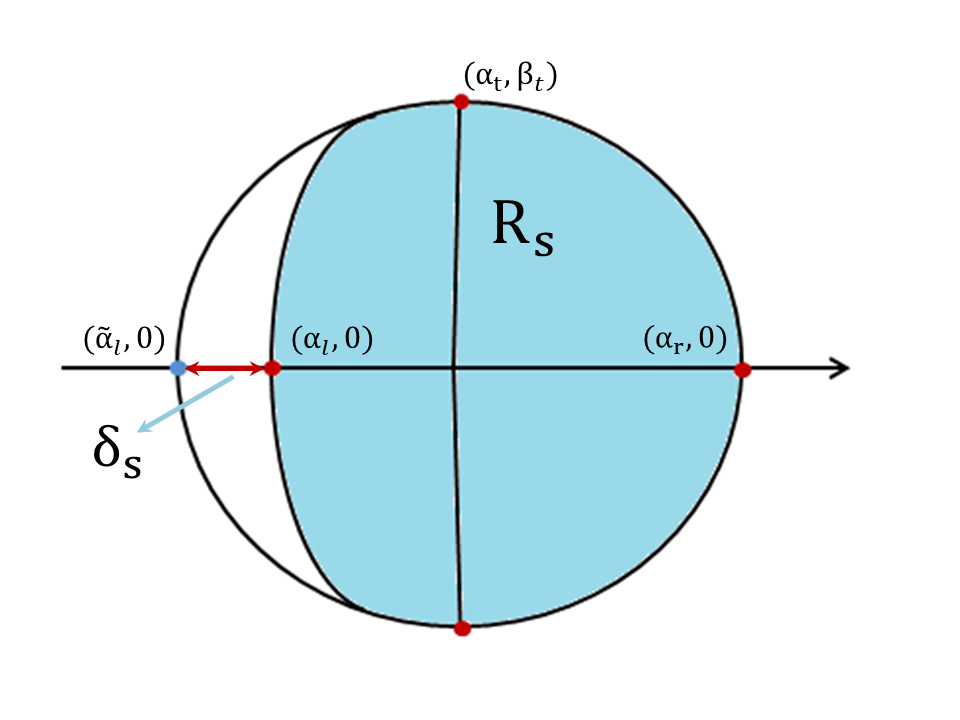}\hspace{-0.3cm}
	\includegraphics[width=0.27\textwidth]{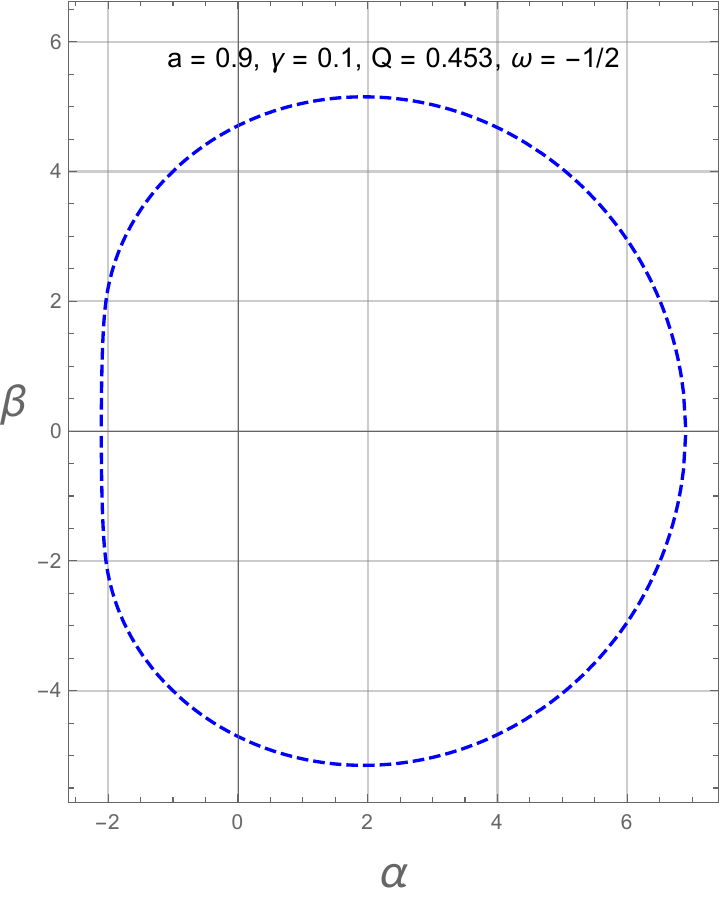}
	\caption{Graphical interpretation of BH shadow and the corresponding reference circle (left), whereas the distorted shadow is cast by a rotating charged BH in the quintessential field (right).}\label{fds}
\end{figure*}
To visualize the visible structure of a spinning charged BH in the
quintessential field, we explain two observables, namely the silhouette radius $\text{R}_s$ and the deformation parameter $\delta_s$~\cite{Hioki:2009na}.
In fact, the shadow radius $\text{R}_s$ can be computed as
\begin{figure*}[ht!]
	\centering	\includegraphics[width=0.4\textwidth]{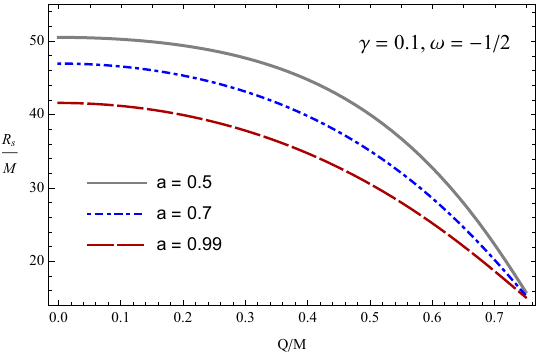}
	\includegraphics[width=0.4\textwidth]{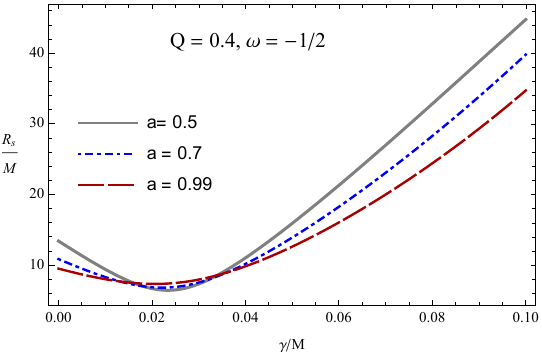}	\includegraphics[width=0.4\textwidth]{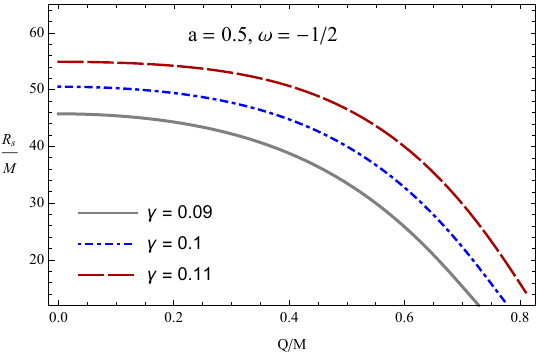}
	\includegraphics[width=0.4\textwidth]{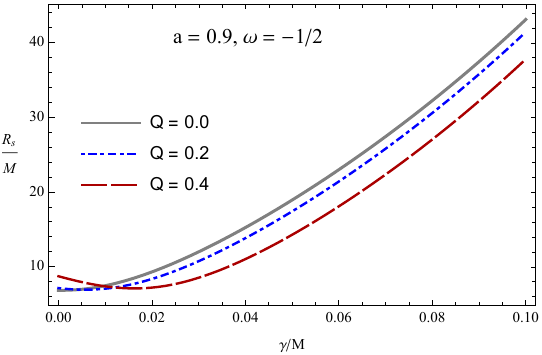}
	\includegraphics[width=0.4\textwidth]{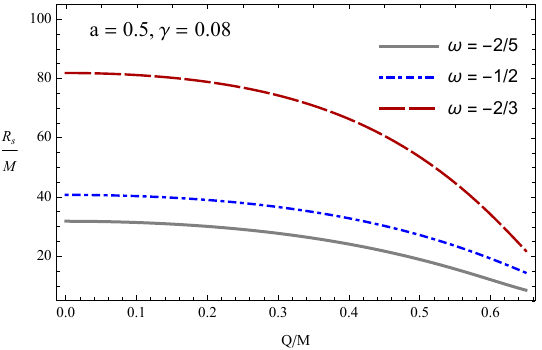}
	\includegraphics[width=0.4\textwidth]{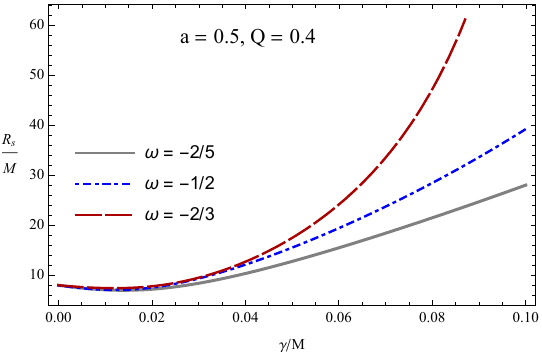}
	\caption{Plots describing the behaviors of Shadow radius $R_s$.}\label{Observable1}
\end{figure*}
\begin{eqnarray}\nonumber
R_s^2 &= & \alpha^2+\beta^2 = \frac{1}{\left( \gamma +2 (M-r) r^{3 \omega } -3 \gamma \omega \right)^2} \Big[ \left( a \gamma  (1-3 \omega )+ 2 a ( M + r) r^{3 \omega }\right)^2+ 2 r \Bigm\{ 4 r^{6 \omega } (-3 M^2 r+2 M Q^2+r^3)\\\label{Rs}
&& + 4 \gamma  r^{3 \omega } (3 M r (\omega-1)+Q^2 (1-3 \omega )) +3 \gamma^2 r (\omega +1) (3 \omega -1)\Bigm\} \Big]\,.
\end{eqnarray}
For non-rotating case, i.e.,~for $a=0$, the above expression simplifies to
\begin{eqnarray}\nonumber
R_s ^2 &= & \frac{1}{(\gamma +2 (M-r) r^{3 \omega }-3 \gamma  \omega)^2}  \Big[2 r \Bigm\{4 r^{6 \omega } (2 M Q^2+r^3-3 M^2 r) +4 \gamma  r^{3 \omega } (3 M r (\omega -1)+Q^2 (1-3 \omega ))\\\label{Rs1}
&& +3 \gamma ^2 r (\omega +1) (3 \omega -1)\Bigm\} \Big].
\end{eqnarray}
While assuming $a = Q = \gamma=0$, the above expression~\eqref{Rs} reduces to the
case of Schwarzschild BH and takes the
form~\cite{1966MNRAS.131..463S,Abdujabbarov:2016hnw}
\begin{equation}
	R_s ^2= \frac{2 r^2 \left(r^2-3 M^2\right)}{(r-M)^2} \,.
\end{equation}
\begin{figure*}[ht!]
	\centering
	\includegraphics[width=0.42\textwidth]{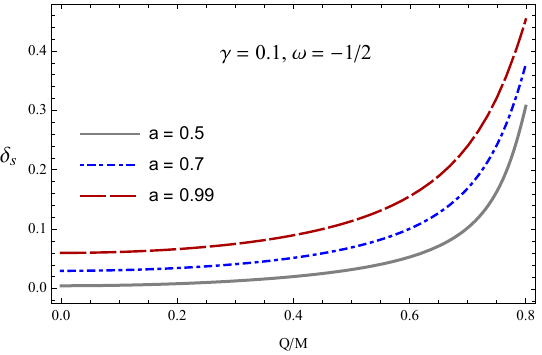}
	\includegraphics[width=0.42\textwidth]{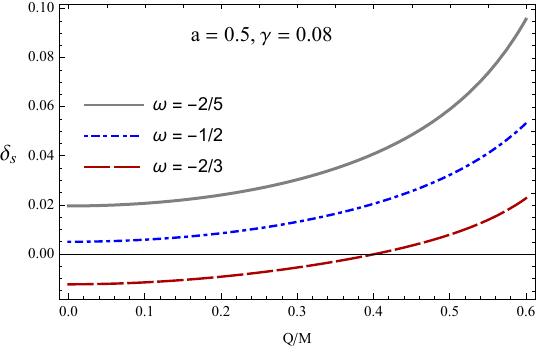}	
	\includegraphics[width=0.42\textwidth]{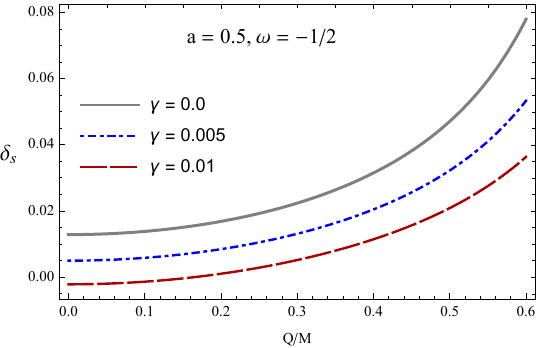}
	\includegraphics[width=0.42\textwidth]{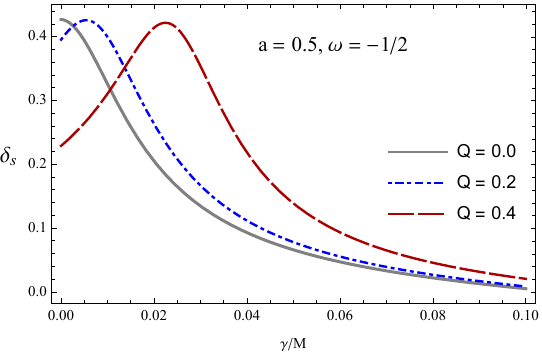}
	\caption{Plots describing the behaviors of deformation parameter $\delta_s$.}\label{Observable2}
\end{figure*}
On the other hand, the distortion parameter $\delta_s$ defines the deviation of the shadow cast by BHs from an ideal circle, which can mathematically be approximated as (see Fig.~\ref{fds}),
\begin{equation}
	\label{deviation} 
	\delta_s=\frac{\mid \tilde{\alpha_l}-\alpha_l \mid}{R_s}.
\end{equation}
In the above expressions, ($\tilde{\alpha_l},0$) and ($\alpha_l,0$) represent the points at which the reference circle and shadow intersect the leftmost $\alpha$ axes. 

The shadow radius $\text{R}_s$ and the distortion parameter $\delta_s$ are plotted in Figs.~\ref{Observable1} and \ref{Observable2}. The shadow radius is plotted in Fig~\ref{Observable1} against charge $Q/M$ (left panels) whereas $\gamma/M$ (right panels). The graphical description of $\text{R}_s$ shows that the quintessence field parameter $\gamma$ contributes to the shadow radius. On the other hand, BH spin $a$, charges $Q$ and $\omega$, reduces the shadow radius.
Similarly, Fig.~\ref{Observable2} illustrates the distortion parameter $\delta_s$. From its graphical interpretation, we observed that $\gamma$ diminishes the distortion effect in the shadow, whereas the parameters $a$, $Q$, and $\omega$ contribute to the distortion effects in rotating BHs shadow.
\section{Constraints from EHT observations}\label{sec:Constraints}
\begin{figure}[ht!]
\centering 
	\includegraphics[width=0.43\textwidth]{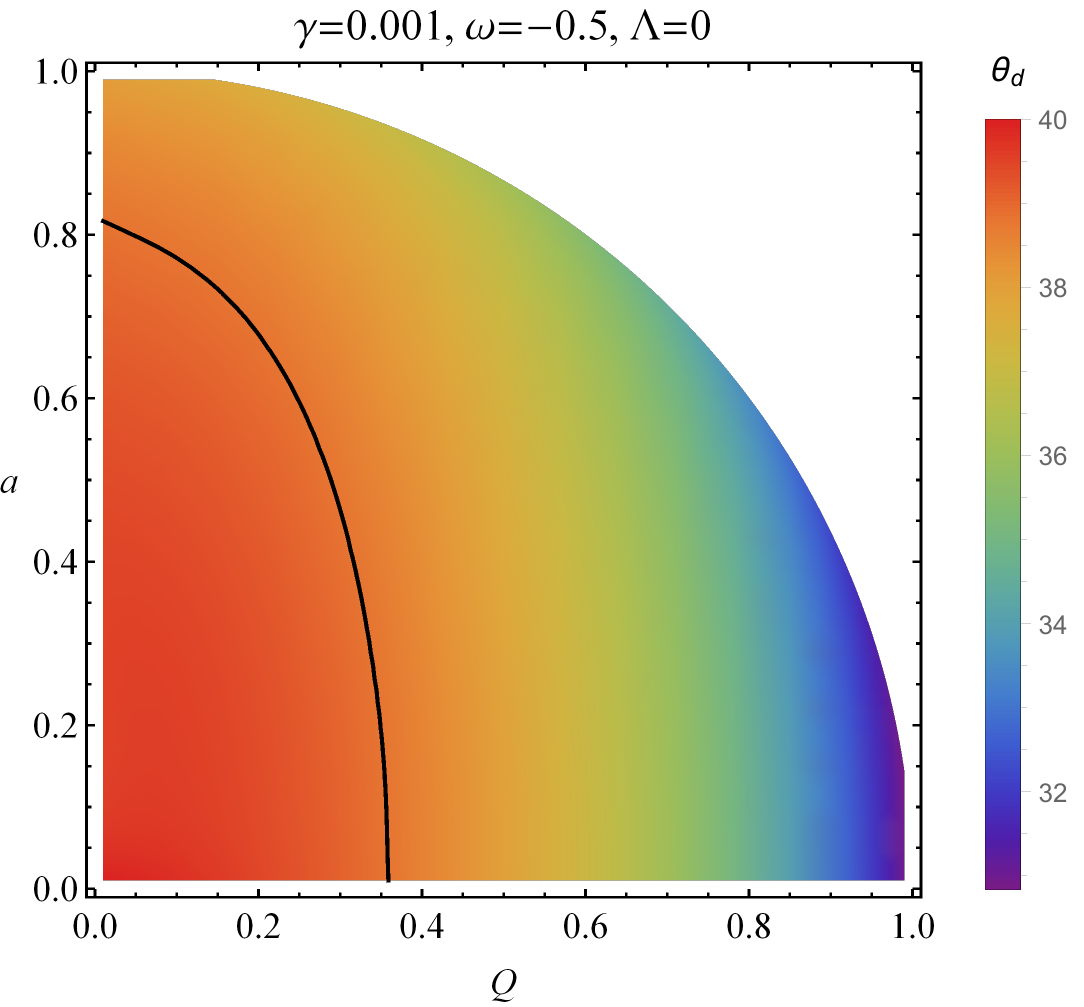}
	\includegraphics[width=0.43\textwidth]{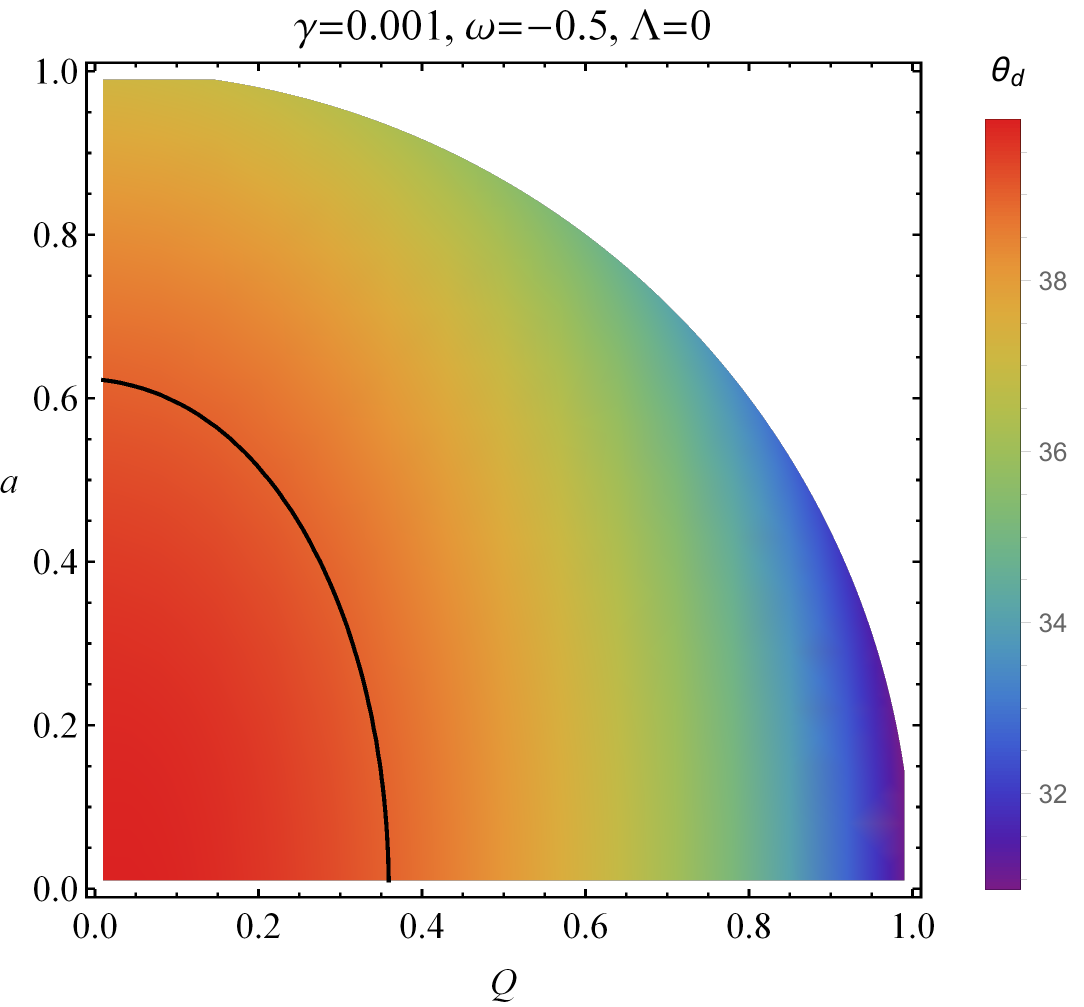}	
	\includegraphics[width=0.43\textwidth]{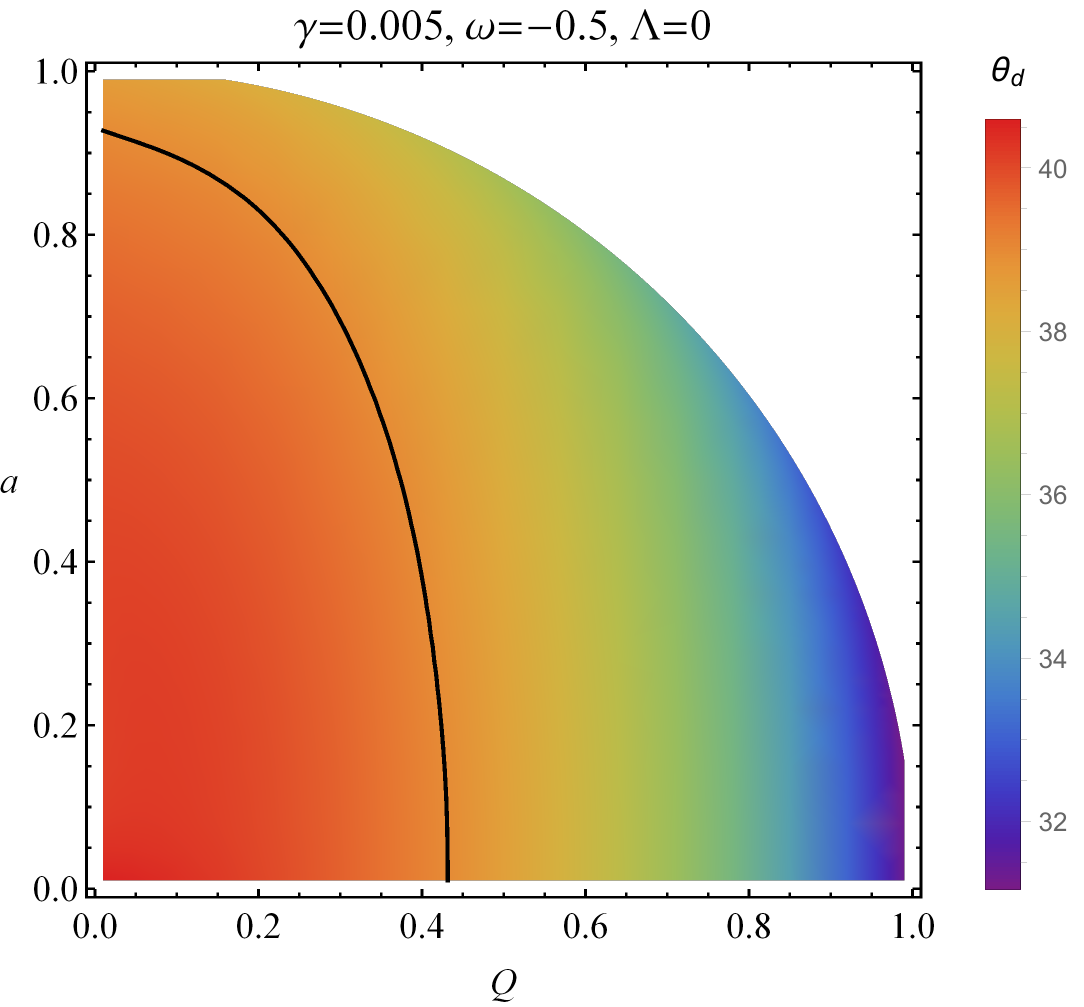}
	\includegraphics[width=0.43\textwidth]{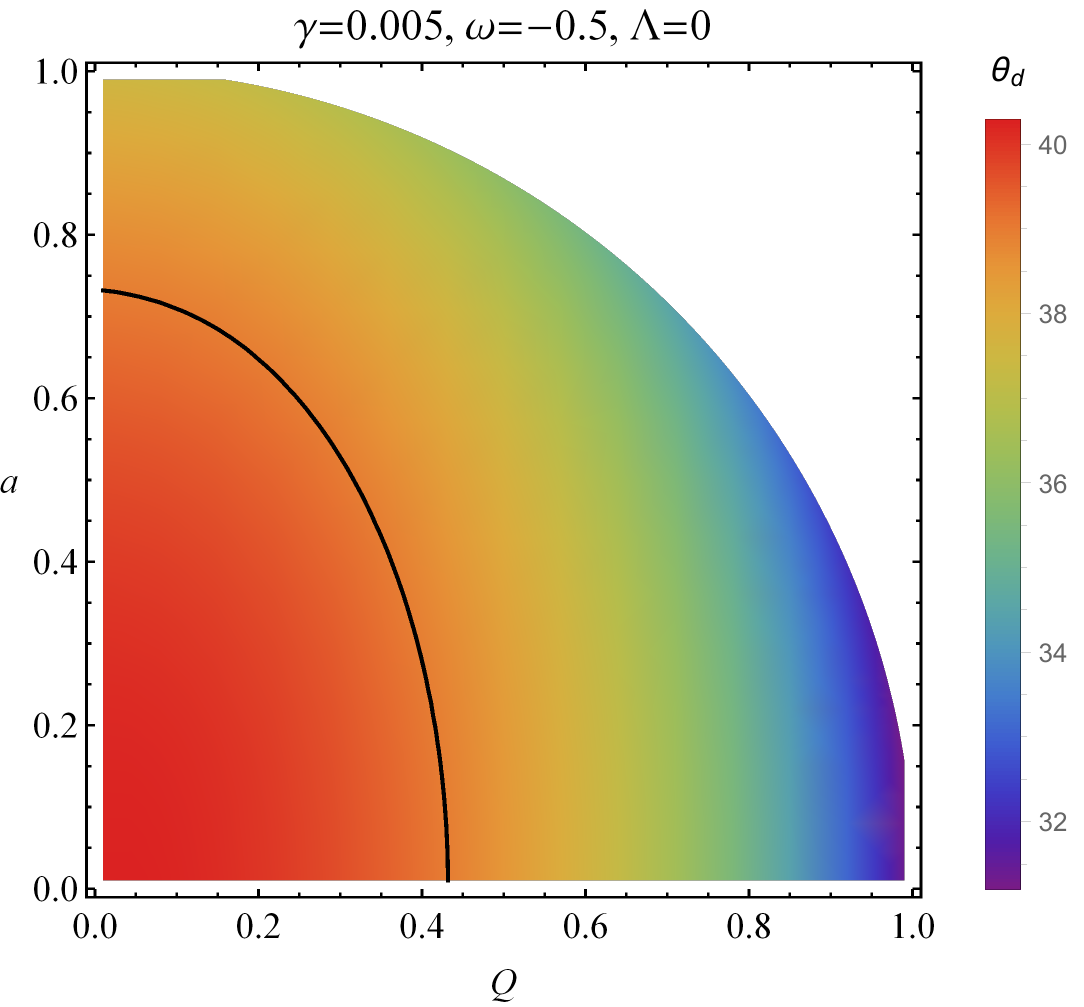}	
\caption{The constraints for M87$^{*}$ for different fixed values of the BH parameters, the inclination angle is {90{$^\circ$}} for the left panels and {17{$^\circ$}} for the right panels. The black curves describe the lower borders ($39~\mu {as}$) of the image size $\theta_d=42\pm3~\mu {as}$ measured angular diameter of the M87* reported by the EHT.}\label{constrM}
\end{figure}

\begin{figure}[ht!]
	\centering\includegraphics[width=0.4\textwidth]{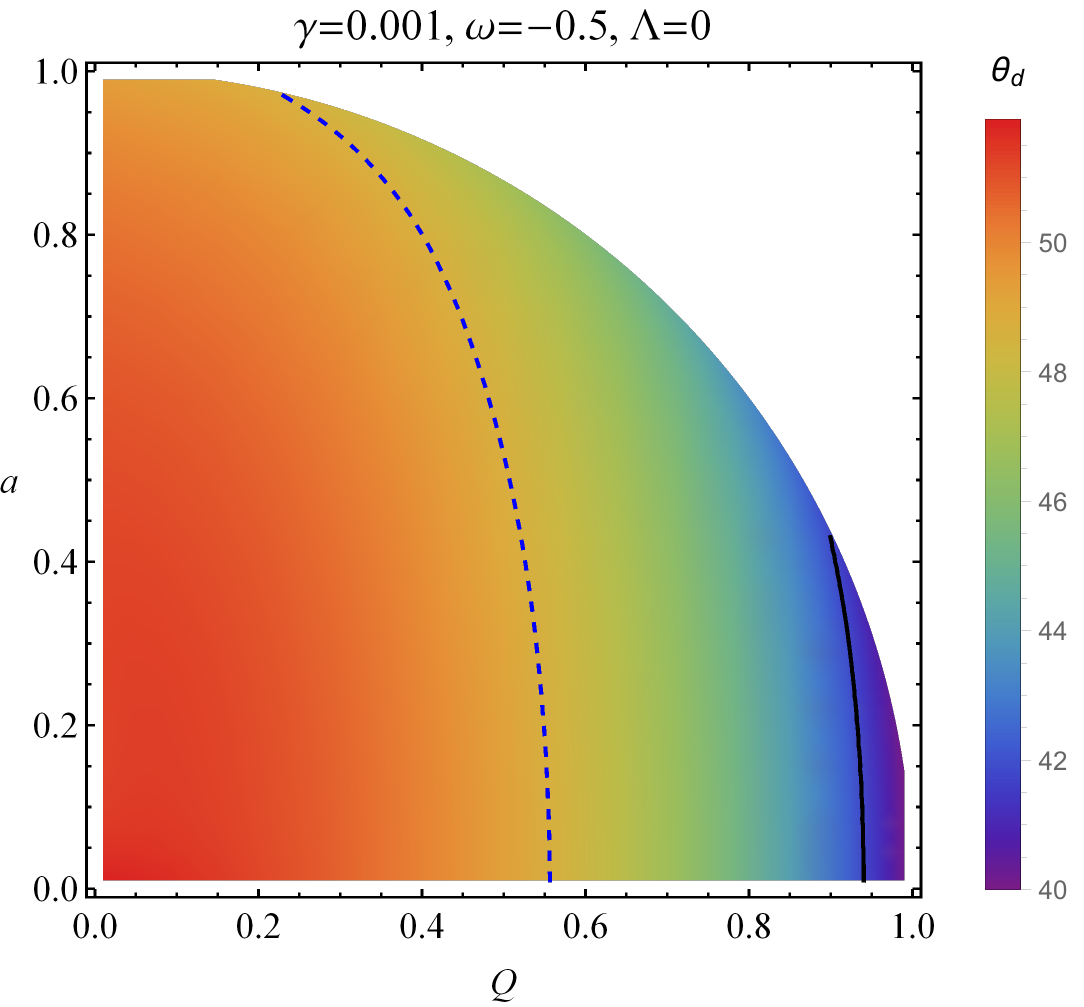}
	\includegraphics[width=0.4\textwidth]{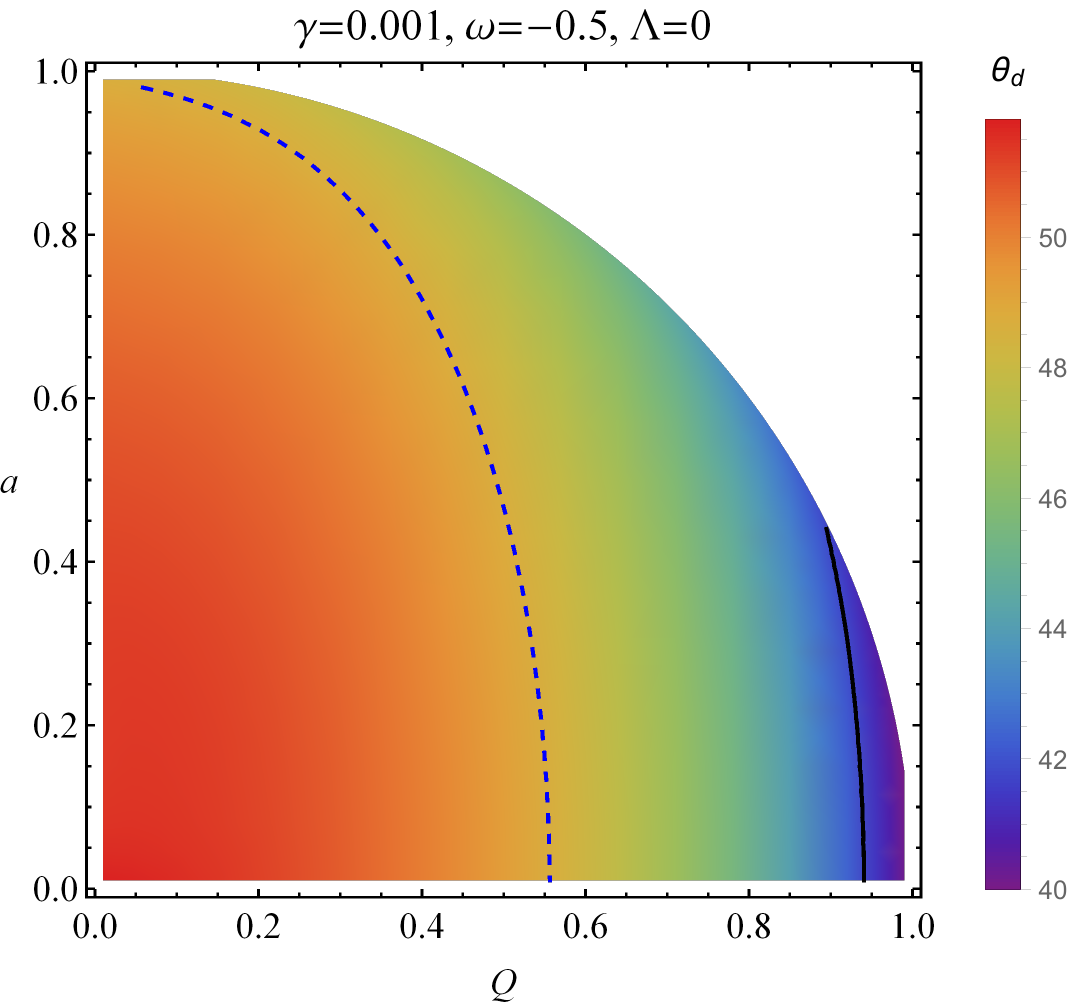}	
	\includegraphics[width=0.4\textwidth]{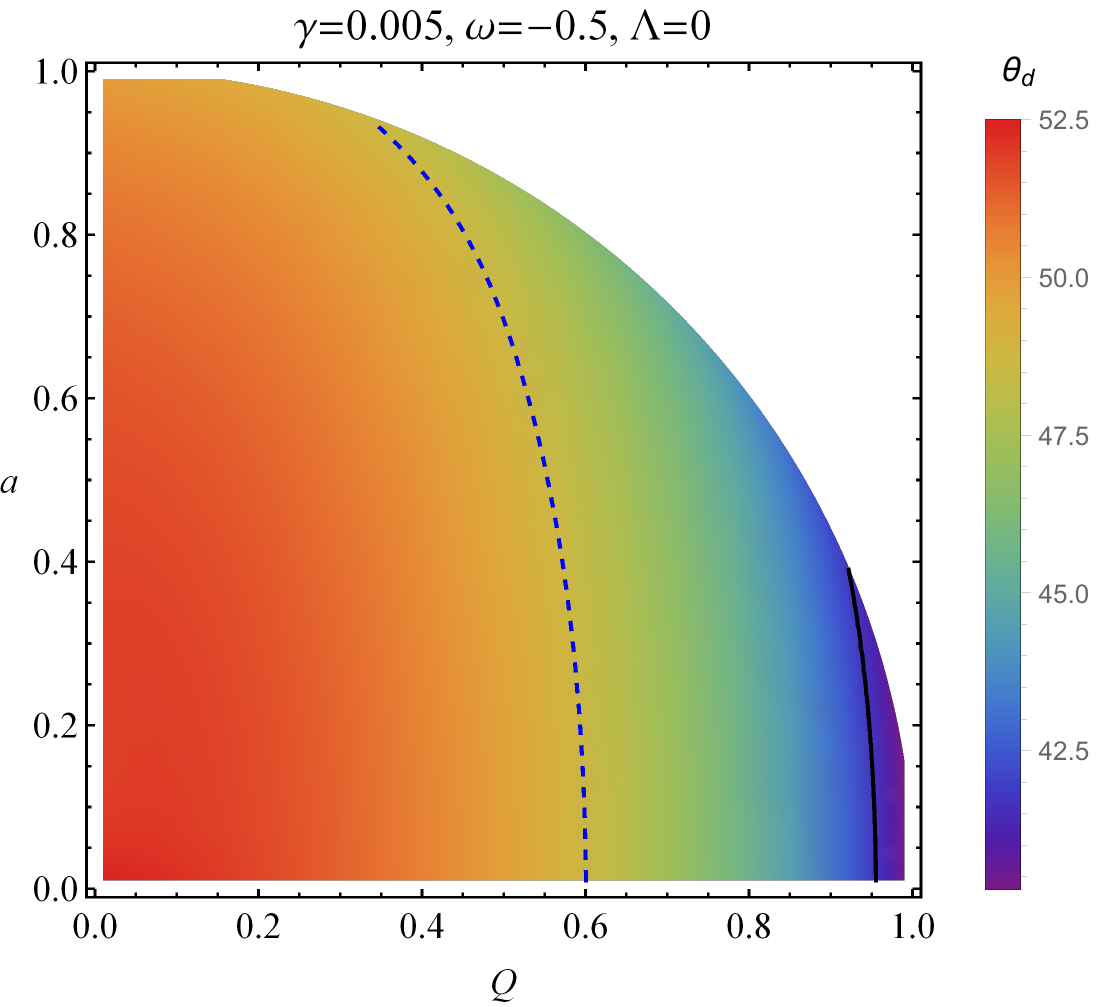}
	\includegraphics[width=0.4\textwidth]{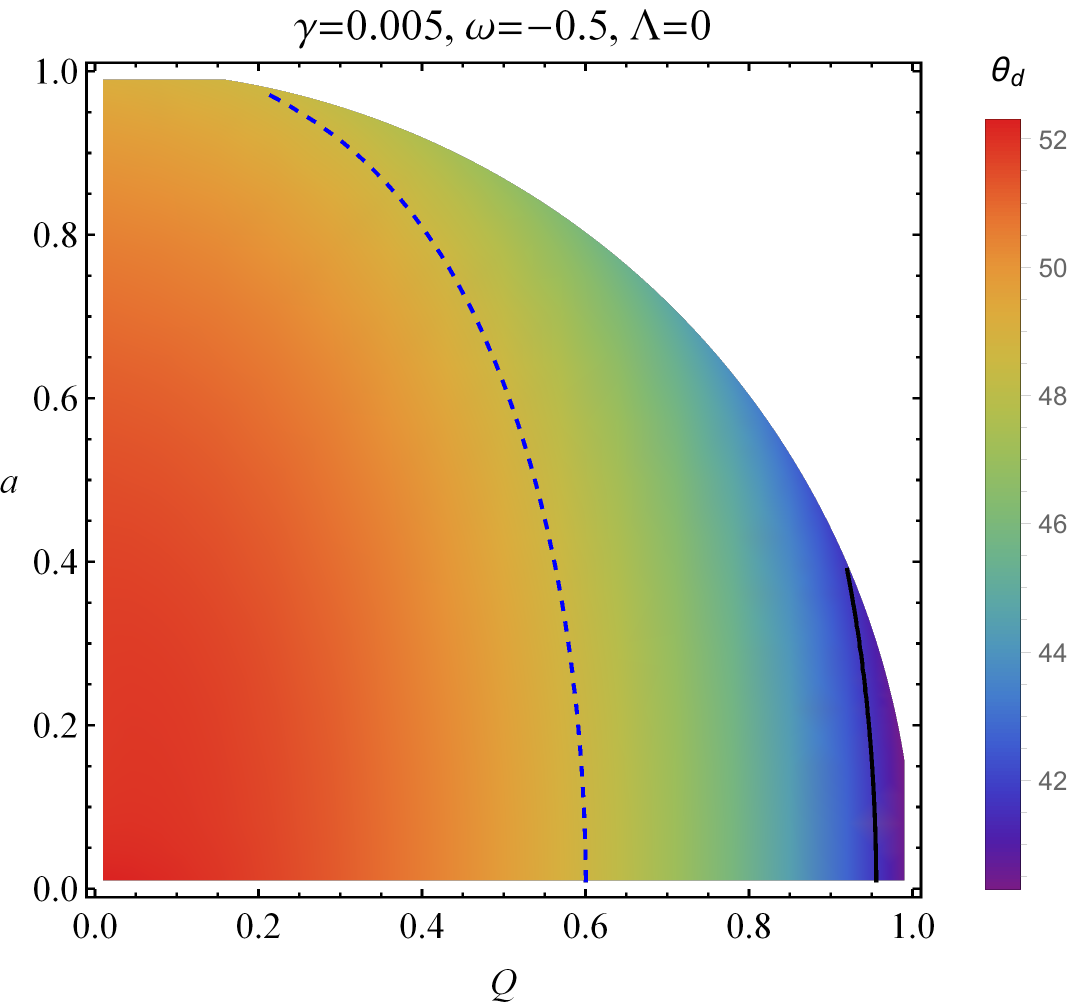}
	\caption{The constraints for Sgr A$^{*}$ are that the angle of inclination is {90{$^\circ$}} (left) and {50{$^\circ$}} (right). The black curves correspond to $\theta_d=41.7~\mu {as}$ within the measured angular diameter, $\theta_d=48.7\pm7~\mu {as}$ of Sgr A$^{*}$ BH reported by the EHT. The blue dashed curves correspond to $48.7~\mu {as}$.}\label{constrS}
\end{figure}

So far, we have only theoretically investigated the shadow of the KNdSQ BH. Still, a natural question arises: Can the shadow of this BH be compared with shadow images of M87$^{*}$ and Sgr A$^{*}$ obtained by the EHT collaboration? In this section, we aim to get constraints from EHT data. Since many modified/alternative gravity models other than general relativity have been proposed, we should check the viability of all these models by comparing them with our observational data.  

To get the constraints, one first needs to define the observable parameters
like the radius and the deviation, which we used to describe the size and shape
of the BH shadow in the previous section.  However, instead of using the Hioki
\& Maedia method, one can express the area of the shadow with a basic geometric
formula in the celestial coordinates and then define two observables--the area
$A$ and the distortion $D$ of the shadow in the following
way~\cite{Kumar:2018ple,Afrin:2021imp}
\begin{eqnarray}
A&=&2\int_{r_-}^{r_+}\biggl(\beta(r)\frac{d\alpha(r)}{dr}\biggr)dr, \label{eq:Area} \\
D&=&\frac{\Delta\alpha}{\Delta\beta}, \label{eq:oblat}
\end{eqnarray} 
where $r_\pm$ refers to the prograde and retrograde radii of circular
orbits obtained as the roots of equation $\eta_c=0$, outside the event
horizon~\cite{Teo:2020sey}.
The Eqs.~\eqref{eq:Area} and \eqref{eq:oblat} can be used to estimate the BH
parameters within the framework of different gravity models by describing the
BH parameters as functions of the observable parameters of the
shadow~\cite{Kumar:2018ple,Afrin:2021imp,Sarikulov:2022atq,Afrin:2022ztr}. 

The angular diameter of the shadow image that an observer measures at a distance $d$ from the BH can be expressed as
\begin{equation}
	\theta_d =\frac{2R_a}{d}, \qquad R_a^2 = {\frac{A}{\pi}}, \label{angdia}
\end{equation}
where $R_a$ is the areal shadow radius. We have finished with the formulations and can get results by using the information about the shadow of M87$^{*}$ and Sgr A$^{*}$ supermassive BHs reported by the EHT collaboration. 

\begin{figure}[ht!]
	\centering\includegraphics[width=0.4\textwidth]{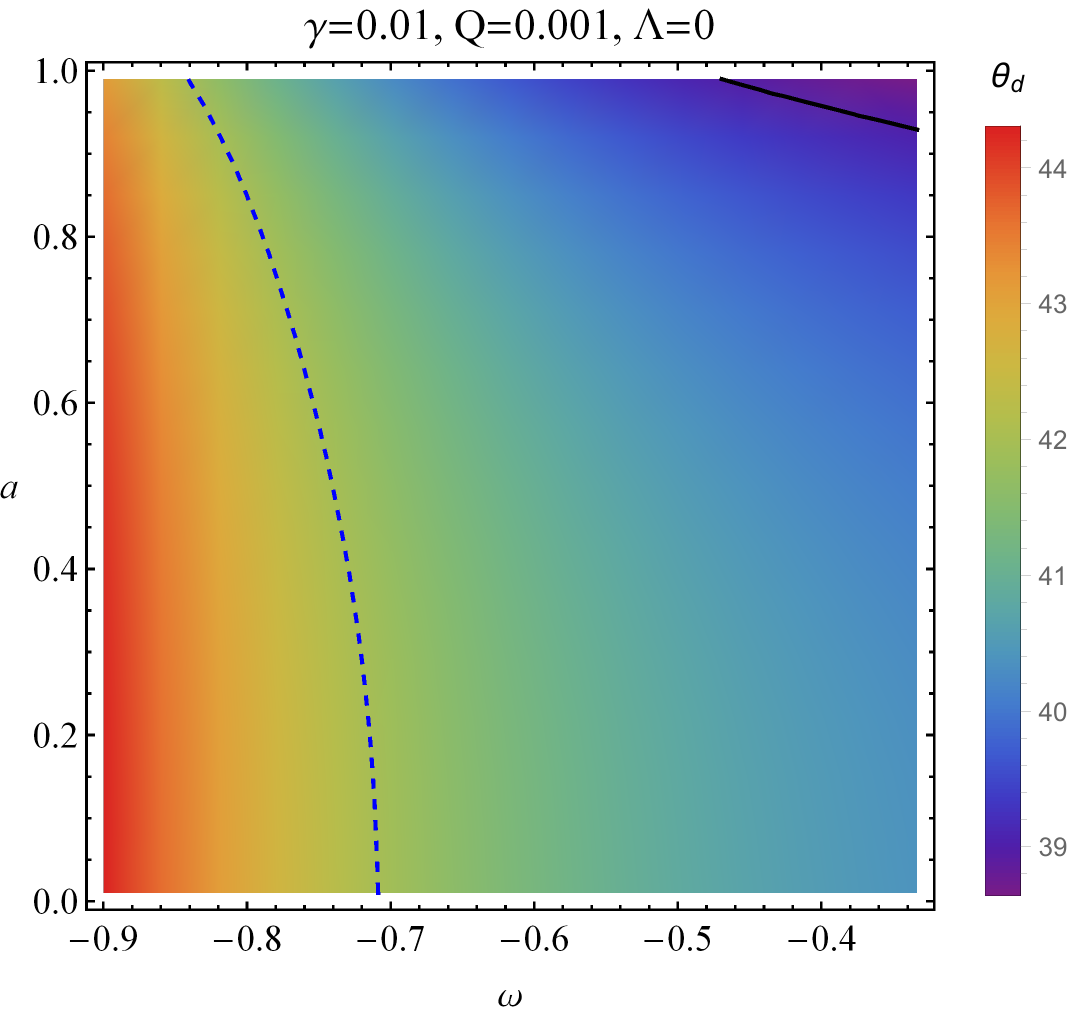}
	\includegraphics[width=0.4\textwidth]{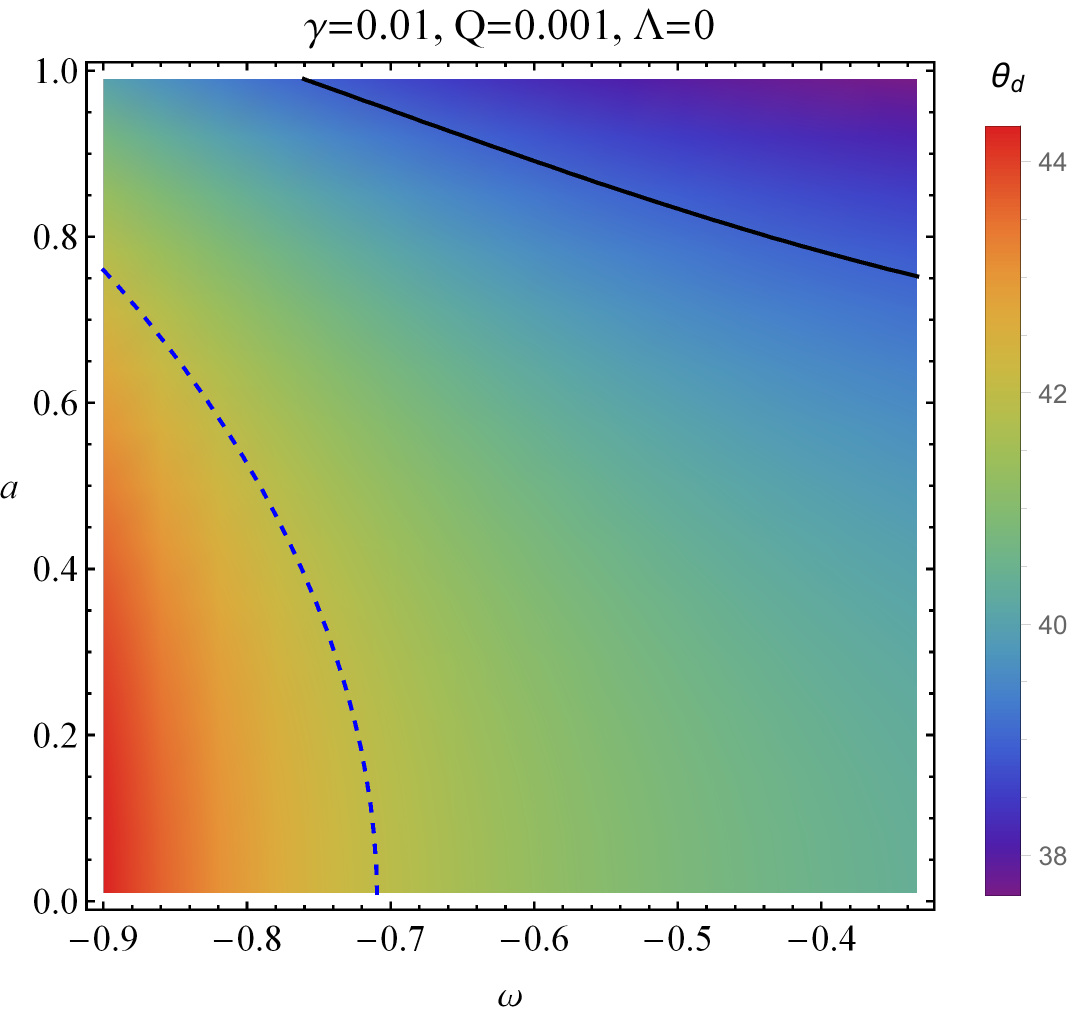}
	\caption{The constraints for M87$^{*}$: angular diameter observable $\theta_d$ for the BH shadows as a function of parameters $a$ and $\omega$ at inclinations {90{$^\circ$}} (left row) and {17{$^\circ$}} (right row). The black and blue dashed curves correspond to $\theta_d=39~\mu {as}$ and $\theta_d=42~\mu {as}$, respectively.}\label{constw1}
\end{figure}
\begin{figure}[ht!]
	\centering\includegraphics[width=0.4\textwidth]{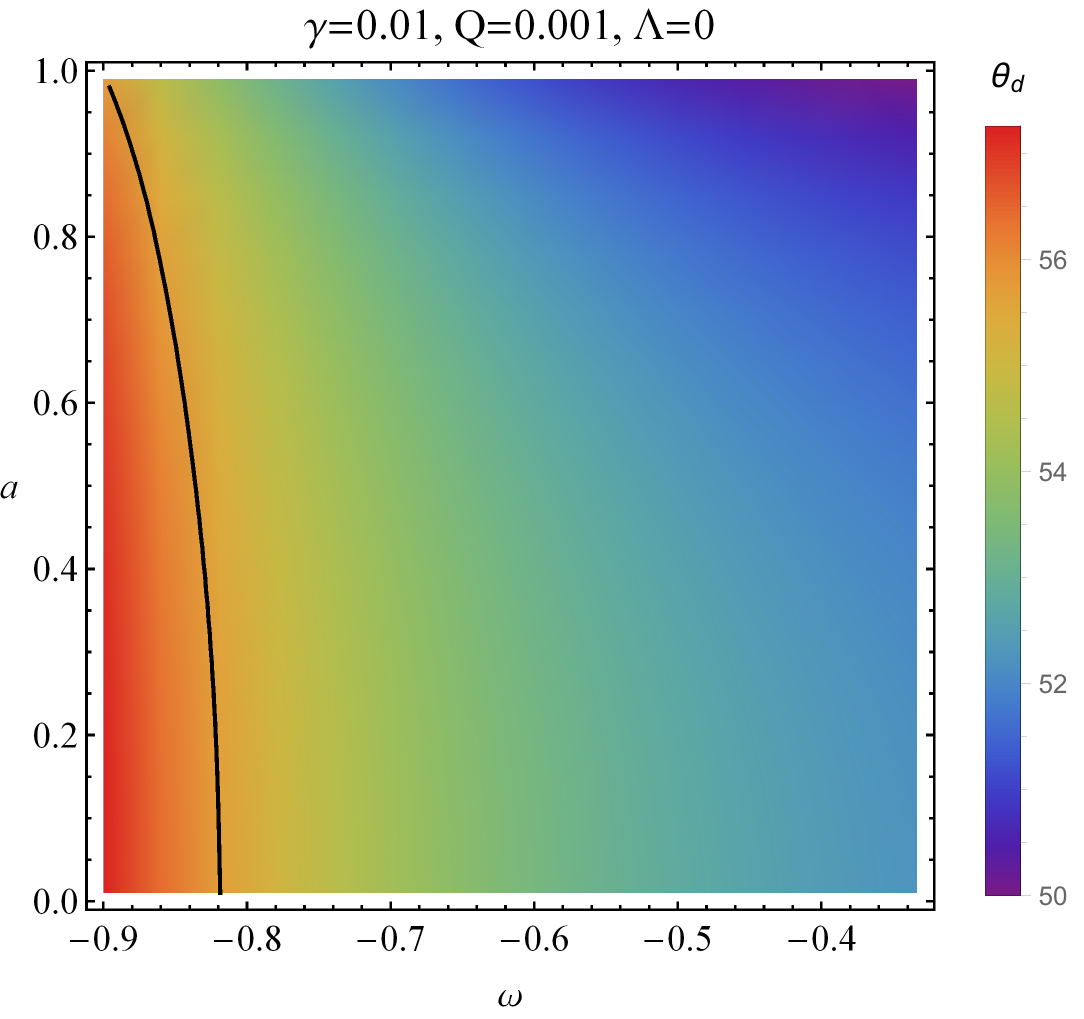}
	\includegraphics[width=0.4\textwidth]{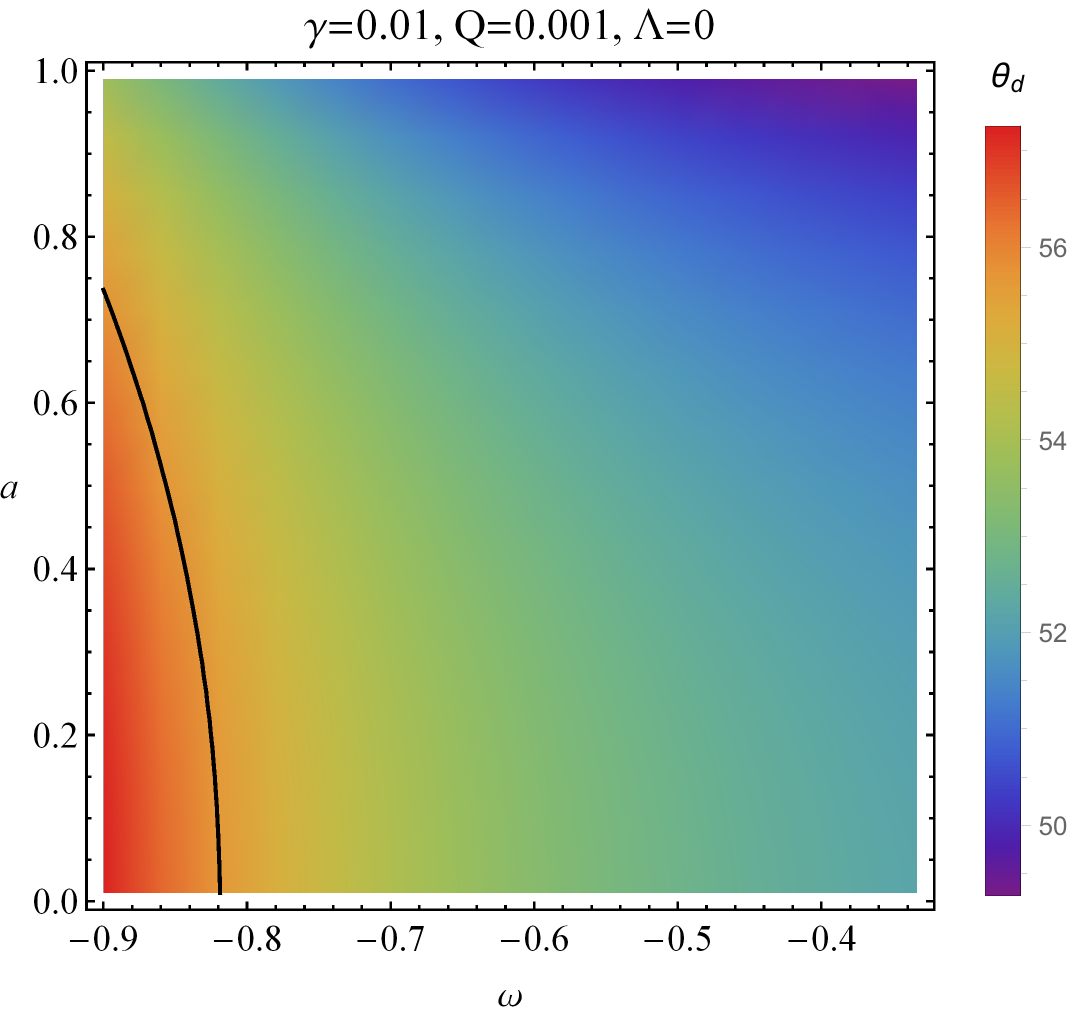}
	\caption{The constraints for Sgr A$^{*}$: angular diameter observable $\theta_d$ for the BH shadows as a function of parameters $a$ and $\omega$ at inclinations {90{$^\circ$}} (left row) and {50{$^\circ$}} (right row). The black curves correspond to $\theta_d=55.7~\mu {as}$.\label{constw2}}
\end{figure}
The angular diameter of the image of the BH M87$^{*}$ is
$\theta_d = 42\pm3 ~\mu as$~\cite{EHT:2019dse}. The mass of M87$^{*}$ and the distance
from Earth can be considered as $M=6.5\times 10^9 M_\odot$ and $d=16.8$ Mpc,
respectively~\cite{EventHorizonTelescope:2019pgp,EventHorizonTelescope:2019ggy}.
For simplicity, in our further calculations, we do not consider the error of
the mass and distance measurements of the BHs located at the center of
M87$^{*}$ and Sgr A$^{*}$. Using Eq.~\eqref{angdia}, one can
express the angular diameter depending on the BH parameters (including mass)
and the inclination angle of the shadow image for an observer located at
infinity. Now, we obtain the constraint values for the supermassive BH
M87$^{*}$ assuming that the BH is a KNdSQ BH. The results are illustrated
in Fig.~\ref{constrM}, the angular size of the shadow on the spin and charge
parameters for the various values of the other BH parameters at inclination
angles {90{$^\circ$}} and {17{$^\circ$}} in left and right panels, respectively.
Here, the black curves correspond to the lower limit of the measured value of
the angular diameter of M87$^{*}$ ($39~\mu as$). It is observed from
the figures that the boundaries of the constraints mainly depend on how the BH
parameters change the size of the shadow since the angular diameter is
proportional to the area radius of the shadow. 
For example, the rise of $\gamma$ causes an increase in the fitted region. Note that change in the inclination angle influences the angular diameter distribution in the $a$-$Q$ space. Moreover, one can observe from Tables~\ref{tab1} and \ref{tab2} that $a$ and $Q$ are inversely proportional, whereas $a$ and $\omega $ are in direct proportion to each other. In other words, in the case of fast-rotating BHs, the constraint value of charge and $\omega$ decreases.
\begin{table*}[ht!]\caption{Numerical results from the $a$-$Q$ space constraints at $\gamma=0.001$, $\omega=-0.5$ and $\Lambda=0$.}
	\label{tab1}
\centering
  \begin{tabular}{@{\hskip 0.5cm}c @{\hskip 0.75cm}c @{\hskip 0.75cm} @{\hskip 2.5cm}c  @{\hskip .75cm}c @{\hskip 0.5cm}}
    \hline \hline \noalign{\smallskip\smallskip}
   \multicolumn{2}{c}{{\hskip -2.75cm} M87*, $39\mu as$}  & \multicolumn{2}{c}{{\hskip -0.7cm}Sgr A*, $48.7\mu as$}    \\ [0.5ex]
    \hline \noalign{\smallskip}
    a &  Q &  a &  Q   \\ [0.4ex]
    \hline  \noalign{\smallskip\smallskip}
    0.0& 0.3595 &  0.0& 0.5565 \\ [0.4ex]
    0.2& 0.3406 &  0.2& 0.5466 \\ [0.4ex]
    0.4& 0.2756 &  0.4& 0.5159 \\ [0.4ex]
    0.6& 0.0902 &  0.6& 0.4568 \\ [0.4ex]
     \hline
\end{tabular}
\end{table*}
\begin{table*}[ht!]\caption{Numerical results from the $a$-$\omega$ space constraints at $\gamma=0.01$, $Q=0.001$ and $\Lambda=0$.}
	\label{tab2}
\centering
 \begin{tabular}{@{\hskip 0.5cm}c @{\hskip 0.75cm}c @{\hskip 0.75cm} @{\hskip 2.5cm}c  @{\hskip .75cm}c @{\hskip 0.5cm}}
	\hline \hline \noalign{\smallskip\smallskip}
	\multicolumn{2}{c}{{\hskip -2.75cm} M87*, $42\mu as$}  & \multicolumn{2}{c}{{\hskip -0.7cm} Sgr A*, $55.7\mu as$}    \\ [0.5ex]
	\hline \noalign{\smallskip}
     a &  $\omega$ &  a &  $\omega$  \\ [0.4ex]
    \hline \noalign{\smallskip}
    0.0& -0.7094 &  0.0& -0.8185 \\ [0.4ex]
    0.2& -0.7227 &  0.2& -0.8251 \\ [0.4ex]
    0.4& -0.7619 &  0.4& -0.8429 \\ [0.4ex]
    0.6& -0.8268 &  0.6& -0.8726 \\ [0.4ex]
     \hline
\end{tabular}
\end{table*}

Similarly, we calculate the constraints on the supermassive Sgr A$^{*}$
parameters, assuming it is a KNdSQ BH. The angular diameter of the shadow of
the Sgr A$^{*}$ is  $\theta = 48.7\pm7~\mu as$~\cite{EventHorizonTelescope:2022wkp}. The mass of BH and the distance from the solar system are $M\simeq4\times10^6M_\odot$ and $d\simeq 8$ kpc,
respectively~\cite{EventHorizonTelescope:2022exc,EventHorizonTelescope:2022xqj}.
Constraints at the angle of inclination {90{$^\circ$}} (left panels) and
{50{$^\circ$}} (right panels) for Sgr A$^{*}$ are demonstrated in Fig.~\ref{constrS}. Here, the black curves correspond to the lower limit
($41.7~\mu as$) in shadow size, while the blue dashed curves describe its
mean value ($48.7~\mu as$). From all these results, one can get information on
which values of the KNdSQ BH parameters can be compared with the Sgr
A$^{*}$ or M87$^{*}$ BHs. 
Just as in the case of M87$^{*}$, here for Sgr A$^{*}$, we observed that the same behavior as $a$ and $Q$ shows an inverse proportion, while $a$ and $ \omega $ are in direct proportion to each other (for details, please see Tables
\ref{tab1} and \ref{tab2}). Surprisingly, in the case of Sgr A$^{*}$, the constraint values of the charge are much greater, while $\omega$ are smaller compared to the supermassive BH M87$^{*}$.

Furthermore, we considered the charge of BH to be very small and put constraints on the quintessential field parameter. Figs.~\ref{constw1} and \ref{constw2} show the angular diameter observable $\theta_d$ for the BH shadows as a function of parameters $a$ and $\omega$ for supermassive BHs M87$^{*}$ and Sgr A$^{*}$. Moreover, our analysis reveals that the variation in the inclination angle greatly affects the $a$-$\omega$ space distribution of angular diameter.
\section{Energy emission}
\label{sec:eemission}
The production and disintegration of enormous particles near the horizon are called energy emissions. The energy emission rate by rotating BHs is an exciting and challenging phenomenon caused by various factors, including Hawking radiation and accretion.
\begin{figure*}[ht!]
\centering
\includegraphics[width=0.418\textwidth]{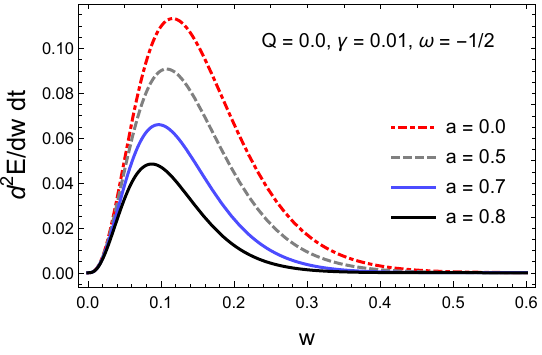}
	\includegraphics[width=0.418\textwidth]{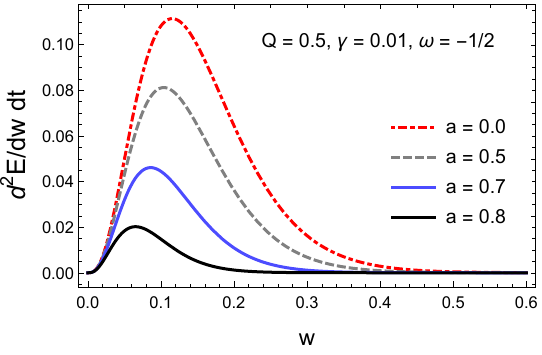}	
	\includegraphics[width=0.418\textwidth]{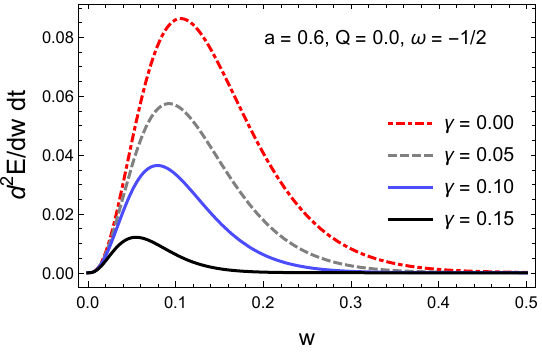}
	\includegraphics[width=0.418\textwidth]{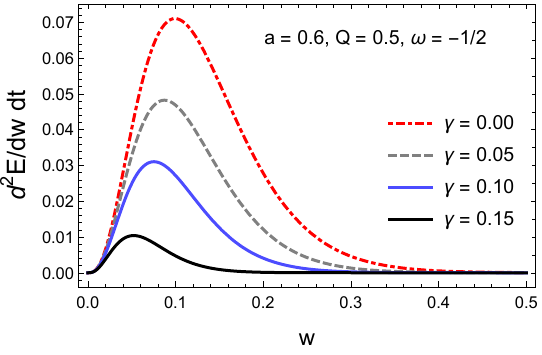}
	\includegraphics[width=0.418\textwidth]{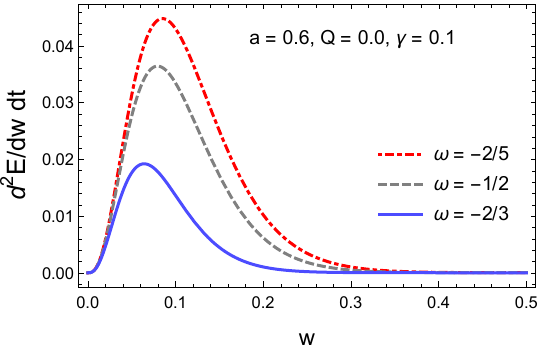}	\includegraphics[width=0.418\textwidth]{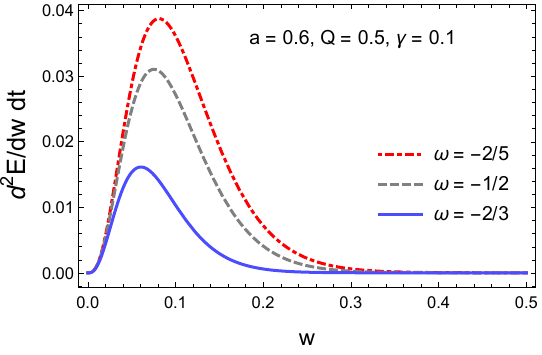}
	\caption{The graphical illustration of energy emission rate along ${w}$, in the left column for non-rotating, while in the right column for rotating cases at various discrete values of the parameters $Q$ and $\gamma$.}\label{En-emission}
\end{figure*}
According to Stephen Hawking's revolutionary theory, BHs emit radiating heat due to quantum events at their event horizons; hence, they are not entirely black. Hawking radiation is a phenomenon that causes BHs to lose mass and energy over time. Generally, a small rotating BH emits higher amounts of radiation and loses energy faster than larger ones, as the energy emission rate through Hawking radiation is inversely proportional to the BH's mass. This has a significant impact on BH's lifespan and final results.

According to our assumption, for a distant observer, BH's shadow reaches the
high-energy absorption cross-section of the BH~\cite{Wei:2013kza}. At extreme
energies, the absorption cross-section's limiting constant value oscillates
around a threshold $\sigma_{ilm}$ (constant value). The threshold of $\sigma_{ilm}$
corresponds to the event horizon radius~\cite{2023FrASS..1074029J}
\[
\sigma_{ilm} \approx \pi R_s^2.
\]
Thus, the energy emission rate can be computed by making use of the limiting constant value as follows,
\begin{equation}\label{ee1}
	\frac{d^2E({w})}{d{w} dt}= \frac{2\pi^2 R_s^2}{\exp^{{w}/T-1}}.
\end{equation}
Here ${w}$ denotes the photon's frequency, while $T$ is the BH temperature at the event horizon, which can be defined as
\begin{eqnarray}\nonumber
	T(r_+)&& = \lim_{\theta\to 0, r\to r_+} \frac{\partial r \sqrt{g_{tt}}}{2\pi \sqrt{g_{rr}}}\\\label{ee2}
	&& = \frac{1}{4 \pi r^{3 \omega}
		\left(a^2+r^2\right)^2} \Bigm[\gamma  \Bigm(a^2 (3 \omega -1) + r^2 (3 \omega +1)\Bigm)-2 r^{3 \omega } \left(a^2 M+r \left(Q^2-M r\right)\right)\Bigm] \,. 
\end{eqnarray}

Fig.~\ref{En-emission} illustrates the behavior of the energy emission rate at
various numerical values of the BH parameters. The left panels show the
charge-less situation ($Q = 0$), while the right panels show the charged
BHs scenario ($Q \neq 0$). We observed that BH charge considerably affects
the energy emission rate. Our results confirm that $\gamma > 0$ diminishes the
energy emission rate, which is consistent with the theory that larger BHs emit
less energy than smaller ones. On the other hand, BH spins $a$ also
diminishes the energy emission rate, which is consistent with the findings
of~\cite{2022ChJPh.78.141K}. We also discovered that raising the absolute value
of the state parameter $\omega$ reduces the rate of energy emission (see
the third row of Fig.~\ref{En-emission}). In other words, $\omega $ contributes
to the BH size; as a result, it results in the decrease of energy emission
rate.

\section{Conclusion}\label{sec:conclusion}
The key objective of this article is to provide a theoretical analysis of the photon region, BH shadow, and its constraints from the EHT observations. {We anticipate our research will be useful for future EHT, James Webb Space Telescope, Square Kilometer Array, and accretion disc observations in various astrophysical environments. Moreover, our findings could provide advancements in the theoretical framework governing quintessential charged BHs.}
First, we have analyzed the horizon structure of the spacetime geometry around
a KNdSQ BH: the horizons and shapes of the apparent regions of the photon
region. We discovered that spin and charge parameters significantly impact the
area of the ergoregion and the photon region, {supporting previous findings~\cite{Khan2019PDU,Fathi:2020agx}}.
Moreover, our acquired results reveal that fast-rotating charged BHs have
greater ergoregion and photon regions than those of chargeless and slowly
rotating BHs. Interestingly, we have also observed that $\Lambda < 0$
increases the photon region, while $\Lambda > 0$ diminishes it, { showing good agreement with the results of~\cite{Chen:2021wqh}}.
It is also observed that $\gamma$ results in shrinking the area of the photon region while contributing to the radii of the event horizon and SLS. 

Next, using the Hamilton--Jacobi formalism, we determined the essential
equations of motion for photons. The apparent shape of a BH shadow can
preferably be visualized with the help of celestial coordinates. Therefore, to
explore the shadow cast by KNSdQ BH, we first obtained the corresponding
celestial coordinates using null geodesics by following Hioki \&
Maeda~\cite{Hioki:2009na}. It has been observed that, due to the effect of the
frame-dragging of spacetime, the BH spin stretches its shadow in the direction
of the rotational axis. Both charge and spin cause distortion in the BH shadow,
consistent with the earlier discovery~\cite{2018EPJC...78}.
Similarly, the quintessence parameter $\gamma$ elongates the shadow radius, while the quintessence field parameter $\gamma$ reduces the distortion effect, which becomes negligible at higher values.
In addition to other parameters, the absolute value of $\omega$ also causes the BH shadow to elongate to the right. In summarizing the effects of DE energy on BH shadow, we can say that DE expands the apparent shape of BH shadow.

We have obtained constraints on BH charge and spin for different values of the quintessential field parameters using the Kumar \& Ghosh technique, where we have calculated the area of BH shadow with a basic geometric formula in the celestial coordinates. 

We have analyzed the angular size of the shadow on the spin and charge parameters for various choices of the BH's other parameters at inclination angles of {90{$^\circ$}} and {17{$^\circ$}}. The boundary values of the constraints are shown to depend on the effect of the BH parameters on the shadow size, which is proportional to the area of the shadow. We have shown that the growth of $\gamma$ causes an increase in the fitted region in the $a$-$Q$ space constraints. Additionally, we assumed that the charge of BH is very small and obtained limitations on the parameters $a$ and $\omega$.

Our analysis of the energy emission rate shows that BH spins $a$, charge $Q$, and $\gamma > 0$ diminish the energy emission rate, {which is consistent with our previous findings~\cite{2022ChJPh.78.141K}, as well as with the theory that states that larger BHs emit less energy as compared to the smaller ones}. In addition, we have also found that raising the absolute value of the state parameter $\omega$ reduces the rate of energy emission (see the third row of Fig.~\ref{En-emission}).
\subsection*{Conflicts of Interest}
The authors declare that they have no known competing financial interests or personal relationships that could have appeared to influence the work reported in this paper.
	
	\def\prc{Phys. Rev. C}
	\def\pre{Phys. Rev. E}
	\def\prd{Phys. Rev. D}
	\def\prl{Physical Review Letters}
	\def\jcap{Journal of Cosmology and Astroparticle Physics}
	\def\apss{Astrophysics and Space Science}
	\def\mnras{Monthly Notices of the Royal Astronomical Society}
	\def\apj{The Astrophysical Journal}
	\def\aap{Astronomy and Astrophysics}
	\def\actaa{Acta Astronomica}
	\def\pasj{Publications of the Astronomical Society of Japan}
	\def\apjl{Astrophysical Journal Letters}
	\def\pasa{Publications Astronomical Society of Australia}
	\def\nat{Nature}
	\def\physrep{Physics Reports}
	\def\araa{Annual Review of Astronomy and Astrophysics}
	\def\apjs{The Astrophysical Journal Supplement}
	\def\aapr{The Astronomy and Astrophysics Review}
	\def\procspie{Proceedings of the SPIE}

	\bibliographystyle{unsrt}
	\bibliography{reference}
\end{document}